\documentclass[aoas,preprint]{imsart}

\RequirePackage{amsthm,amsmath,amsfonts,amssymb}
\RequirePackage[authoryear]{natbib}

\startlocaldefs

\usepackage{natbib}
\usepackage{graphicx}%
\usepackage{multirow}%
\usepackage{amsmath,amssymb,amsfonts}%
\usepackage{amsthm}%
\usepackage{mathrsfs}%
\usepackage[title]{appendix}%
\usepackage{xcolor}%
\usepackage{textcomp}%
\usepackage{manyfoot}%
\usepackage{booktabs}%
\usepackage{algorithm}%
\usepackage{algorithmicx}%
\usepackage{listings}%

\usepackage[utf8]{inputenc}

\usepackage{hyperref}       
\usepackage{booktabs}       
\usepackage{amsfonts}       
\usepackage{nicefrac}       

\usepackage{xcolor}         
\usepackage{wrapfig}        

\usepackage{comment}

\usepackage{amssymb}
\usepackage{amsthm}

\usepackage{blindtext}
\usepackage{booktabs}
\usepackage{varwidth}
\usepackage{caption}

\usepackage{dsfont}

\usepackage{graphicx}

\usepackage{paralist}
\usepackage{mathrsfs}
\usepackage{upgreek}
\usepackage{newtxmath} 
\usepackage{cleveref}
\usepackage{multirow} 
\usepackage{enumitem}

\usepackage{placeins}

\newcommand{\lns}{$\log(N)-\log(S)$ }
\newcommand{\lnsn}{$\log(N)-\log(S)$}
\newcommand{\chandra}{{\sl Chandra}}
\newcommand{\indep}{\overset{\text{indep}}{\sim}}

\newcommand{\subregion}{${\mathcal{S} }$}

\newcommand{\name}{HELIOS}

\usepackage{soul,xcolor} 

\usepackage{algorithm}
\usepackage[noend]{algpseudocode}

\usepackage{caption}
\usepackage{subcaption}

\usepackage{tikz}
\usetikzlibrary{shapes, arrows}
\usepackage{capt-of}

\usetikzlibrary{arrows.meta}
\tikzset{>={Latex[width=2mm,length=2mm]}}

\usepackage{url} 

\usepackage[multiple]{footmisc}

\usepackage{mathtools}
\usepackage{amsmath}
\usepackage{cases}

\usepackage{adjustbox}

\theoremstyle{thmstyleone}%

\endlocaldefs

\begin{document}

\begin{frontmatter}
\title{
Combining Multiple Incomplete Astronomical Surveys to Estimate Galaxy Flux Distributions
}
\runtitle{Flux Distributions from Multiple Incomplete Surveys}

\begin{aug}
\author[A]{\fnms{Maximilian}~\snm{Autenrieth}\ead[label=e1]{ma2216@cam.ac.uk}\orcid{0009-0006-2068-5950}},  
\author[B]{\fnms{David  A.}~\snm{van Dyk}\ead[label=e2]{d.van-dyk@imperial.ac.uk}},
\author[C]{\fnms{Vinay}~\snm{Kashyap}\ead[label=e3]{vkashyap@cfa.harvard.edu}},
\author[D]{\fnms{Andreas}~\snm{Zezas}\ead[label=e4]{azezas@physics.uoc.gr}},
\author[C]{\fnms{Aneta}~\snm{Siemiginowska}\ead[label=e5]{asiemiginowska@cfa.harvard.edu}},
\and
\author[E]{\fnms{David C.}~\snm{Stenning}\ead[label=e6]{dstennin@sfu.ca}}
\address[A]{Department of Pure Mathematics and Mathematical Statistics,
University of Cambridge
\printead[presep={ ,\ }]{e1}}

\address[B]{Department of Mathematics,
Imperial College London \printead[presep={ ,\ }]{e2}}
\address[C]{Center for Astrophysics,
Harvard \& Smithsonian \printead[presep={ ,\ }]{e3,e5}}
\address[D]{Department of Physics,
University of Crete \printead[presep={ ,\ }]{e4}}
\address[E]{Department of Statistics and Actuarial Science,
Simon Fraser University \printead[presep={ ,\ }]{e6}}
\end{aug}

\begin{abstract}
In astrophysics, the distributions of fluxes or intensities of populations of sources provide insight into the structure and evolution of the universe.
We propose a science-driven hierarchical Bayesian framework to estimate the X-ray flux distribution of a population of galaxies. The distribution is parameterized via a broken power law model, formulated as a mixture of Pareto distributions, and is fit using data from a combination of non-representative X-ray and optical surveys. 
Because fainter sources are less likely to be detected, astronomical surveys suffer from incompleteness at the low-intensity end, and dedicated survey simulations are used to study the resulting missing-data pattern and determine the form of corrections. 
We introduce a methodology to combine data from two (or more) surveys with different incompleteness functions. This allows us to leverage deeper surveys (e.g., optical) to more fully characterize the X-ray \lnsn\ (where fewer sources are detected).  
We take advantage of a well-calibrated characterization of the missing-data pattern, particularly in X-ray surveys, to make incompleteness corrections for sources that are not detected in either survey. 
The proposed model is validated using a simulation study designed to mimic the Chandra Deep Field South (CDF-S), demonstrating accurate recovery of the flux distribution even under high proportions of systematic incompleteness. Finally, we apply our proposed method to data from the CDF-S.
\end{abstract}


\end{frontmatter}


\section{Introduction}\label{sec:introduction_LNS}

The flux distributions of populations of astronomical sources provide clues as to the geometry and structure of the universe, e.g., aiding our understanding of  stellar and galactic evolution \citep{scheuer1957statistical,hewish1961extrapolation,mathiesen1998constraints,guetta2005constraining,mateos2008high,tranin2024statistical}.  
Thus, estimating these distributions is of key importance for effectively all areas of astrophysics. However, their determination is subject to several biases, selection effects, and  other complications. 
The luminosity of a source is its intrinsic energy output per unit time, while its flux is the energy per unit time per unit area measured by a detector at a given distance from the source. 
Thus, the flux distribution of a population of sources does not account for differences in source distance.
Historically, the flux distribution of a population is characterized by $N( >S)$, the number of sources with flux greater than $S$. Astronomers are particularly interested in the relationship between $\log_{10}(N(> S))$ and $\log_{10}(S)$, in short \lnsn. In the literature, the \lns relationship is commonly assumed to be linear or piecewise linear \citep[e.g.,][]{murdoch1973maximum,georgakakis2008new}, allowing different flux ranges to capture changes in the composition or evolution of the source populations. 

In this paper, we focus on estimating the X-ray \lnsn\ function for populations of galaxies. X-rays are high-energy photons emitted by astronomical objects such as galaxy clusters, stars, supernova remnants, neutron stars, and accreting black holes, including those in X-ray binary systems.  
Since X-rays are mostly absorbed by the Earth's atmosphere, X-ray surveys are obtained via space-based telescopes, such as the {\sl Chandra X-ray Observatory}. X-ray telescopes count individual photons, and source fluxes are estimated using the observed cumulative photon energies (per unit time and area). A simpler measure is the source intensity, $\lambda$, i.e., the expected rate of source photon counts (per unit time). 
The expected count rate of a source can be converted to its flux using the expected photon energy distribution specific to the source, calculated from its fitted spectrum \citep{arnaud2011handbook}. 
When studying a population that is homogeneous in terms of its source spectra, as in this paper, the source fluxes are roughly proportional to the count rates and can be computed via a detector-specific ``count-to-flux'' conversion.

\subsection{Statistical challenges and current approaches}\label{section:Challenges_and_prospects}

Generally, the \lns distributions are determined from observations of areas of the sky in the spectral regions of interest (e.g., optical, infrared, X-ray etc.) followed by the detection of the sources and measurement of their intensity and flux.
As a result, the analysis of the \lns\ distributions is subject to several biases related to the limitations of the observations, selection biases, and statistical effects.   All of these effects need to be accounted for in order to infer the correct underlying \lns distribution of an observed source population. To demonstrate these effects and our proposed method, we consider the \lns distribution of galaxies observed in deep X-ray surveys.
The Chandra Deep Field South (CDF-S) is the longest exposure obtained with \chandra\ \citep{xue2011chandra,luo2017chandra}. It consists of 52 individual \chandra\ ACIS-I exposures, obtained over more than 10 years with a total exposure of approximately 44 days. Here, we use the CDF-S to illustrate the statistical challenges of estimating \lnsn, including background contamination of the source counts and incompleteness of the detected sources.  

The true source location combined with the instrumental point spread function (PSF; the spread of source counts due to the finite spatial resolution of the telescope) determines the region of the X-ray detector that records the photons associated with each source. The measured counts in each such region are a mixture of source and background X-ray photons. A na\"ive, indeed possibly negative, estimate of the source intensity simply subtracts an estimate of the background count rate (obtained from a putative source-free region of the detector) from the overall count rate in the source region. 
Figure~\ref{figure:logN-logS_XrayMle} illustrates an estimated \lns curve obtained via these na\"ive estimates (brown crosses) using detected X-ray sources from a subregion of the CDF-S. Specifically, Figure~\ref{figure:logN-logS_XrayMle} plots one minus the empirical cumulative distribution function (CDF) of the na\"ive estimates, renormalized to the number of detected X-ray sources.  Unfortunately, this estimate exhibits severe bias owing to incompleteness of the detected sources, i.e., low-intensity sources are far less likely to be detected than those with high count rates. The detection thresholds of standard methods \citep[e.g.,][]{freeman2002wavelet} are understandably designed to limit false detections and thus inevitably reduce the power to detect faint sources. This effect is exacerbated by the Poisson nature of X-ray counts whereby sources with the same count rate can have highly variable and skewed observed counts, particularly at low intensities. 
This leads to a bias in the sample of detected sources relative to the true intensity distribution of the population, as upward fluctuations of the numerous faint sources are more likely to cross the detection threshold than downward fluctuations of brighter sources are to be excluded. This effect is known as Eddington bias \citep{eddington1913formula,teerikorpi2004influence}. 
At the same time, brighter and nearer objects are more likely to be recorded in astronomical surveys (e.g., due to detection sensitivity limits of the telescopes), leading to non-representative survey data, an effect commonly referred to as Malmquist bias \citep{malmquist1920study}.
These effects are combined with telescope and survey-specific properties, such as varying exposure times, differential blurring of sources across the detector, etc. Overall, these effects form a complex \textit{non-ignorable} missing data mechanism \citep{little2019statistical} with the potential to significantly bias statistical results. 
 
Many classical methods for estimating \lnsn\ functions assume complete surveys or incorporate incompleteness only through simplified corrections. 
Assuming Gaussian measurement errors for the estimated fluxes, for example, \cite{crawford1970maximum} and  \cite{murdoch1973maximum}, 
fit a linear \lnsn\ via
maximum likelihood (ML). 
\cite{maccacaro1982medium} and
\cite{schmitt1986number} extend this framework, again modeling a linear relationship, but replacing Gaussian approximations with a Poisson model for the errors, which is more appropriate for low photon counts. 
\cite{loredo1995inferring} develop a Bayesian approach to gamma-ray burst number–flux distributions that accounts for measurement uncertainty, nondetections, and survey selection effects. 
\cite{georgakakis2008new} extends the linear model to a piecewise linear \lns relationship, also assuming Poisson errors. \cite{zezas2007chandra} incorporates uncertainty in the flux-to-count conversion parameters as part of their ML framework. 
\cite{wong2014automatic} formulate the piecewise linear \lnsn\ as a double (or multiple/broken) Pareto population distribution for intensity parameters within a hierarchical model. This flexible model allows 
fitting the number and locations of breakpoints  
via a novel interwoven EM algorithm.  \cite{wong2014automatic} apply their method to a subset of the Chandra Deep Field North selected to avoid the issue of incompleteness, which, as acknowledged by \cite{wong2014automatic}, ``can lead to tremendously more complicated and challenging modeling." 
Related hierarchical Bayesian approaches have been developed for galaxy luminosity-function estimation under measurement error and flux-dependent selection \citep{szalai2018gpu}. 
A pair of PhD theses \citep{udaltsova2014universe, stampoulis2017bayesian} formulate hierarchical models that explicitly account for survey incompleteness by introducing a large number of unobserved sources. Under Pareto population models, however, the expected number of sources grows rapidly at lower intensities, implying a substantial missing population. A direct missing-data formulation would therefore involve a large fraction of missing information \citep{vand:meng:01art} and lead to inefficient posterior sampling.
To avoid this, both works marginalize over the missing-source variables and instead work with a fixed-dimensional posterior. However, this marginalization introduces likelihood terms involving parameter-dependent integrals over the population and detection models, which must be evaluated repeatedly during MCMC and can become a computational bottleneck. 
These models are formulated for a single survey and do not extend to combining multiple surveys with differing levels of incompleteness.

\subsection{Contribution}\label{section:contribution}

\begin{figure}[t]
\centering
\begin{subfigure}{0.32\linewidth}
    \centering
    \includegraphics[width=\linewidth]{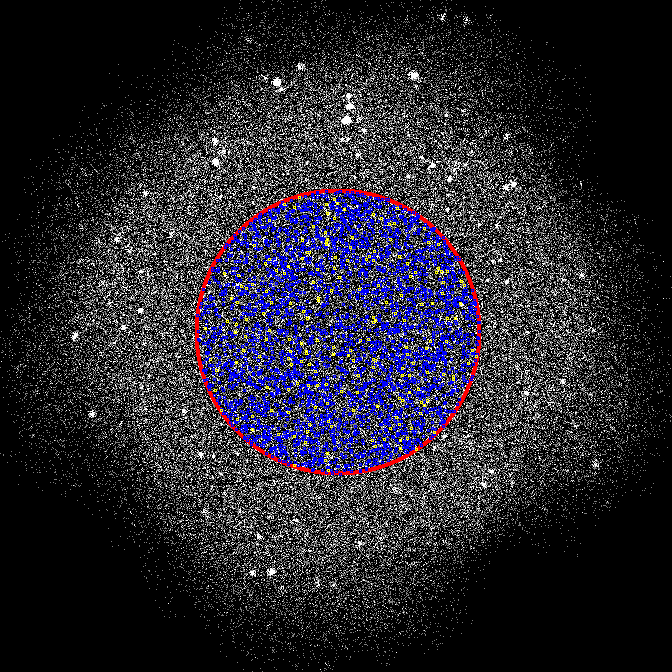}
    \caption{CDF-S event image.}
    \label{figure:CDFS_event_file}
\end{subfigure} 
\begin{subfigure}{0.32\linewidth}
    \centering
  \includegraphics[width=\linewidth]{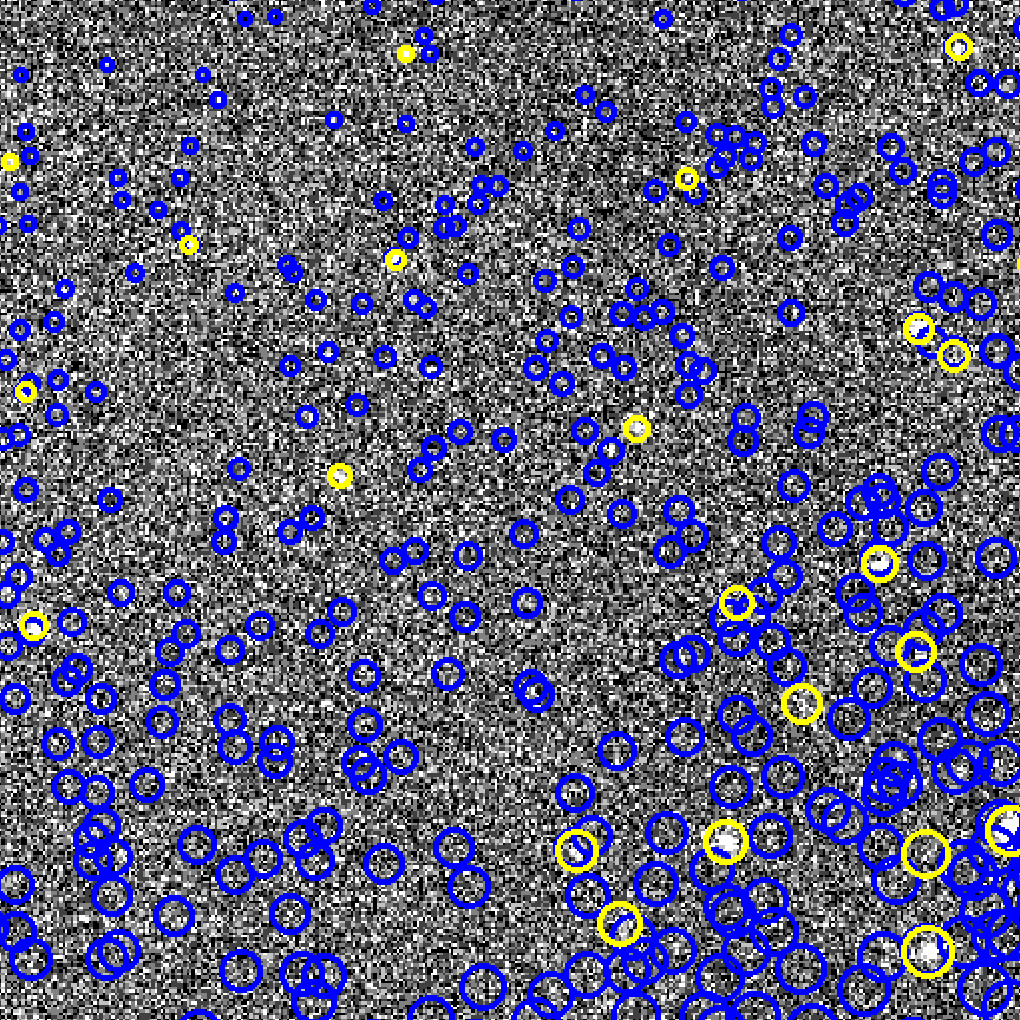}
    \caption{Patch of the CDF-S region.}
    \label{figure:patch_of_CDFS_all_sources}
\end{subfigure}
\begin{subfigure}{0.32\linewidth}
    \centering
    \includegraphics[width=\linewidth]{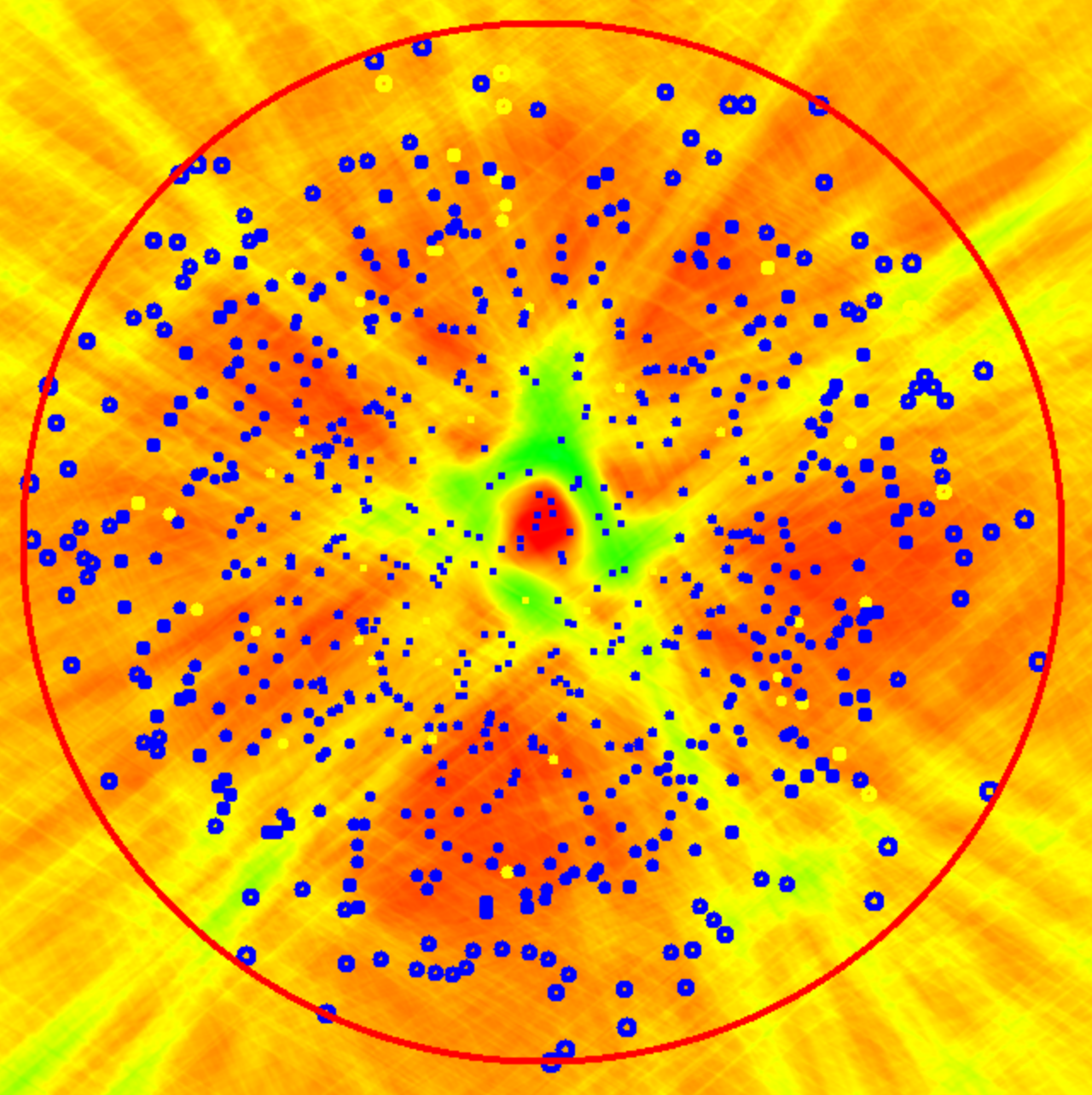}
    \caption{CDF-S Exposure map.}
    \label{figure:exposure_map_main}
\end{subfigure}
  \caption{
(a) CDF-S X-ray event image (merged 2000, 2007, and 2010 Chandra observations). The red
circle  with radius of 5 arcmin indicates the analyzed subregion $\mathcal{S}$ at the center of the
field. 
(b) Zoomed-in patch of $\mathcal{S}$. X-ray detected sources are shown in
yellow and optical sources in blue; photon counts are displayed in grayscale.
Source regions are circular and increase in size with off-axis angle due to
the widening \chandra\ point spread function (PSF).
(c) Exposure map of $\mathcal{S}$. Circles indicate the sources retained in
the final sample (yellow: X-ray detected, blue: optical-only). The exposure
varies across the field, and normalized effective areas $e_i$ are derived from the
exposure map to account for this spatial variation.
  }
  \label{figure:CDFS_data}
\end{figure}

We propose using an external, more complete survey to probe systematic incompleteness of an X-ray survey. This could involve, for example, using an optical, radio, or previous X-ray survey to 
identify a more complete set of sources in the population. In our data analyses, we extract the locations of sources detected in an optical catalog \citep[][details appear in Section~\ref{section:CDFS_analysis}]{arnouts2001eso} within the CDF-S subregion illustrated in
Figure~\ref{figure:CDFS_event_file}. We transfer the optical source locations to the X-ray survey and define their respective source areas via the X-ray PSF. Because the optical survey is much richer than the CDF-S, we expect to associate X-ray photons with a much larger set of candidate sources, particularly low-intensity sources. Figure~\ref{figure:patch_of_CDFS_all_sources} illustrates a patch of the CDF-S subregion with X-ray detected sources identified as yellow circles and sources detected in the optical catalog as blue circles.   
The number of optical sources is much larger than the number of X-ray detected sources.  
Another advantage of this approach is that restricting attention to X-ray sources with optical counterparts mitigates contamination from distinct astrophysical subpopulations and defines a more homogeneous galaxy population for inference.

\begin{figure}
  \centering
  \includegraphics[width=0.7\linewidth]{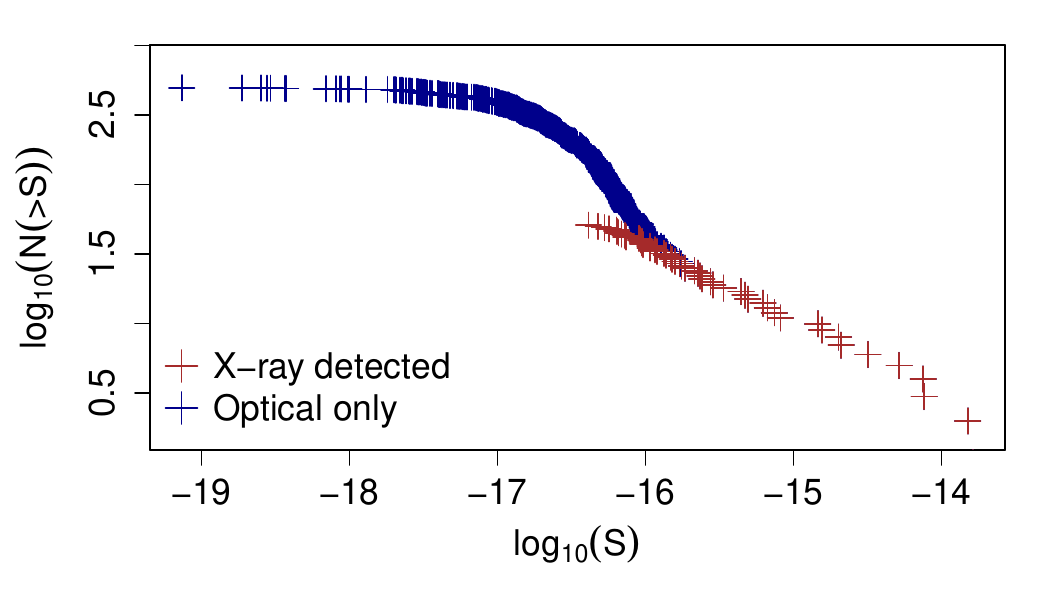} 
\caption{Illustration of a na\"ive $\log(N) - \log(S)$ estimation via eCDF of source intensities obtained by simply subtracting an estimate of the background counts in the source region from the observed counts in that region. The estimated $\log(N) - \log(S)$ curves are shown for sources detected in X-ray (in brown), 
and for sources identified in an optical catalog, with X-ray flux estimated from the corresponding X-ray counts (in blue). 
In both cases the source samples are treated as complete, leading to an apparent discrepancy in the $\log(N) - \log(S)$ estimate and a very strong turnover at low fluxes, illustrating the strong necessity of incompleteness corrections. \label{figure:logN-logS_XrayMle} 
} 
\end{figure}

To illustrate the power of combining an optical and an X-ray survey, we return to the CDF-S example in Figure~\ref{figure:logN-logS_XrayMle}. Specifically, we extract the X-ray counts from source areas obtained by combining the optical locations with the X-ray PSF and obtain a  na\"ive (background subtracted) estimate of the corresponding source intensities. About half of these na\"ive estimates are negative, confirming the need for a more sophisticated estimate. 
Using the empirical CDF of this augmented 
set of intensity estimates, we recompute the \lnsn\ and plot it with blue crosses in Figure~\ref{figure:logN-logS_XrayMle}. (For simplicity, we only use non-overlapping source areas and positive intensity estimates in this demonstration.) Even with poor quality count-rate estimates, the strong discrepancy between the \lns relationships obtained via the X-ray sample and via the X-ray sample augmented with optical sources illustrates the magnitude of incompleteness among the X-ray detected sources. In particular, the optical sources reach much farther into the low-intensity end of the distribution. 

Figure~\ref{figure:patch_of_CDFS_all_sources} illustrates several practical challenges. Matching optical and X-ray sources, for example, is not always straightforward \citep{budavari2015probabilistic}.  Although most of the X-ray sources have a well-identified optical counterpart, that is, an optical source region that contains the center point of the X-ray source region, 
this is not always the case.  Some X-ray sources have no recognizable optical counterpart and others have more than one.   
As expected, a large proportion of optical sources have no detected X-ray counterpart. Inevitably, there are sources that appear in neither the X-ray nor the optical catalog. Finally, there are many overlapping optical source regions. Dealing with overlap is a significant challenge, since the observed counts in the overlapping regions are a mixture of photons from the several sources, which need to be disentangled for accurate estimation of the source-specific intensities \citep{meyer2021ebascs}.\footnote{ 
Detected X-ray source regions are non-overlapping by construction of the X-ray detection and source-extraction procedure, given the instrumental resolution. 
The optical catalog has finer resolution and detects sources 
in closer proximity. However, here the source areas for X-ray and optical sources are calculated via the X-ray PSF, which is much larger than the optical PSF. This leads to significant overlap among optical source regions.}   There are additional technical challenges, such as errors in source locations, incorporating realistic PSFs, instrumental exposure and off-axis corrections, etc., which are discussed  
throughout the paper. The exposure map for the CDF-S subregion in Figure~\ref{figure:CDFS_event_file} is illustrated in Figure~\ref{figure:exposure_map_main}.

To estimate \lnsn\ functions using data from multiple surveys with varying levels of incompleteness, we propose \name\ 
(Hierarchical Estimation of $\text{Log}(N)-\log(S)$ from Incomplete Observational Surveys), a Bayesian hierarchical model for combining incomplete surveys. 
We postulate a broken power-law population distribution for the source intensity parameters, formulated as a mixture of Pareto distributions and corresponding to a piecewise linear \lnsn, a commonly used parametrization in X-ray studies (e.g.,  \citealt{barger2005cosmic,wong2014automatic,ueda2014toward,udaltsova2014universe}). 
Within each source area, photon counts are modeled as an exposure-adjusted sum of independent Poisson background and source counts, thus allowing principled modeling of the background contamination. We account for incompleteness in each survey by incorporating separate incompleteness functions.
In our CDF-S case study, we use an X-ray incompleteness function, known from previous simulations \citep{wright2023simulating}, and model the optical incompleteness function, with parameters learned from the data. In particular, we demonstrate how the known X-ray incompleteness function can be leveraged to learn the optical incompleteness function.

To avoid introducing numerous latent variables representing the population of unobserved sources, 
we formulate our population model in terms of the number of astronomical sources in bins defined by narrow source intensity ranges and the combination of surveys in which they are detected.
This construction builds on the discretization underlying Lindsey’s method \citep{lindsey1974comparison,lindsey1974construction}, in which continuous distributions are represented through bin counts. 
The corresponding Poisson-regression formulation of the binned counts provides a convenient approach to density estimation 
and has subsequently been used and extended in a range of settings, including specially designed exponential families, empirical-Bayes density estimation and Tweedie corrections, and conditional density estimation \citep{efron1996using,efron2004large,efron2008microarrays,efron2011tweedie,gao2022lincde}. 
Here, we build on this Poisson-binning idea in a different setting, where the Poisson structure arises naturally from a random source-population size, survey incompleteness is modeled through intensity-dependent thinning, and the quantities being binned are latent source intensities embedded within a Bayesian hierarchical model. 

More precisely, under this construction, the bin counts are modeled as Poisson random variables with expectations determined by a product of the incompleteness functions or their complements (as determined by the missing-data pattern of the bin), the population distribution (e.g., double Pareto), the expected population size, and the bin width. In the limit as the intensity bins narrow to zero, this Poisson likelihood equals the target likelihood; with larger bins, it serves as a tunable computationally efficient approximation.  
This allows us to deploy a Metropolis-Hastings within Gibbs sampler for our overall hierarchical model, thus obtaining estimates and uncertainty quantification of the parameters of both the \lnsn\ and incompleteness functions. To the best of our knowledge, this results in the first principled approach that combines multiple surveys to estimate the galaxy \lnsn\ in X-rays, while carefully accounting for incompleteness in the respective surveys.

The remainder of the paper is organized as follows. 
In Section~\ref{section:source_model}, we review 
the Bayesian hierarchical model for estimation of \lnsn\ under the assumption of complete surveys. 
In Section~\ref{section:poisson_binning}, which forms the core methodological contribution of the paper, we build on the work of \cite{lindsey1974comparison,lindsey1974construction} and \cite{efron1996using} to develop a general \emph{Poisson-binning} method to handle non-ignorable missing data. First, in Section~\ref{section:binning_one_survey}, we describe the general approach in the context of a single incomplete survey. In Section~\ref{section:binning_multiple_surveys}, we extend the approach to allow a principled combination of multiple incomplete surveys. 
In Section~\ref{section:incompleteness}, we propose our hierarchical Bayesian model for principled estimation of \lnsn\ via Poisson-binning with two (or more) incomplete surveys. In Section~\ref{section:incompleteness_marginalization}, we discuss the X-ray incompleteness function of \cite{wright2023simulating},
and we outline the efficient implementation of the incompleteness corrections in Section~\ref{sec:implement}. 
In Section~\ref{section:CDFS_analysis}, we present a comprehensive case study. We first describe the CDF-S data acquisition and processing in Section~\ref{section:CDFS_data}.
We then validate the proposed methodology via a suite of simulation studies in Section~\ref{section:sim_studyLogNlogS}. These studies are designed to capture key characteristics of the CDF-S data and the incompleteness mechanism. 
In Section~\ref{section:CDFS_posterior}, we apply the model to CDF-S data augmented with the optical catalog of \cite{arnouts2001eso}, presenting posterior inference and the resulting \lns estimate.
We conclude with a discussion of the method and findings in Section~\ref{sec:discussion}.  
Supplementary materials include details of the prior specification, incompleteness corrections, and Metropolis-within-Gibbs sampler, along with additional analyses and diagnostics for the CDF-S data and simulation study.

\section{A complete-data hierarchical model for \lnsn}
\label{section:source_model}

Following the framework of \cite{wong2014automatic}, we begin with a review of the hierarchical source model for \lnsn\ that assumes a complete X-ray survey with no missing sources. Given an observation time of $\mathcal{T}$ (seconds), let $Y_i$ be the photon count in source region $i$, for $i = 1, \dots, n$, 
where $n$ 
is the number of observed sources. Each source region corresponds to a contour of the X-ray PSF and is centered at the source location. Although our framework allows for an arbitrary PSF, in our numerical studies we use a bivariate normal PSF with i.i.d. components so that the source regions are circular; we set the radii to $3\sigma$ of the PSF to capture about $99\%$ of the source photons. In practice, the radius varies across the detector; see Section~\ref{section:CDFS_analysis} for details. The area of source region $i$ is denoted by $a_i$ (pixels).

The observed counts, $\mathbf{Y} = (Y_1, \ldots, Y_n)$, 
are each a sum of independent Poisson background and source counts and are themselves mutually independent. Thus, they are modeled as 
\begin{equation}\label{eq:observed_model}
Y_i \mid \left(\lambda_i, \xi\right) \stackrel{\text{indep}}{\sim} \text{Poisson}  \left((a_i\xi+ r_i e_i \lambda_i)\mathcal{T} \right),
\end{equation}
where $\xi$ is the background count rate (counts/sec/pixel), $\lambda_i$ is the count rate for source $i$ (counts/sec), $e_i$ is the telescope's normalized effective area in the source region, and \( r_i \) is the PSF-determined proportion of photons from the source expected to fall within the source region (in our case $ r_i \equiv 0.99$). We denote the collection of count rates by $\pmb{\lambda} = (\lambda_1, \ldots, \lambda_n)$. 
We assume that background is distributed uniformly across the field of view so that $\xi$ is the same for each source, but allow the effective area to vary across the detector. 
Our goal is to estimate the distribution of the source intensities $\lambda_i$.

We also observe $X$ photons from a background region of area $A$ (pixels), free of detected sources, i.e., 
\begin{equation}
X \mid \xi\sim
 \text{Poisson} (A \xi \mathcal{T}). \label{eq:background_model_uniform}
\end{equation} 
\noindent 
We specify the prior distribution on $\xi$ as $\text{Gamma}(a_{\xi}, b_{\xi})$, where the gamma distribution is parameterized so that its mean equals $a_{\xi} / b_{\xi}$.

Detailed work has been performed to determine the best-fit population distribution for source luminosities / intensities \citep[e.g.,][]{georgakakis2008new,maccacaro1982medium,wong2014automatic,udaltsova2014universe,stampoulis2017bayesian}. 
\cite{udaltsova2014universe} shows that a linear \lns relationship corresponds to a Pareto population distribution (single power law) and a piecewise-linear \lns relationship with $m$ pieces corresponds to an $m$-component broken power law. The latter is formulated as a mixture of $m-1$ truncated Pareto distributions and one untruncated Pareto distribution \citep[Lemma 2,][]{udaltsova2014universe}, i.e.,
\begin{equation}
    p_\Lambda(\lambda \mid \pmb{\theta} ) = \sum_{j=1}^m  
    \left[ \prod_{k=1}^{j-1}  \left(\frac{\tau_{k+1}}{\tau_k}\right)^{-\theta_k}  \right]
    \left(\frac{\theta_j}{\tau_j}\right)\left(\frac{\lambda}{\tau_j}\right)^{-(\theta_j +1)}
    \mathds{1}_{\{ \tau_j \leq \lambda < \tau_{j+1} \} },
    \label{eq:lambda.prior}
\end{equation} 
where $\pmb{\theta}$ denotes the collection of parameters, $\pmb{\theta} = (\theta_1, \ldots, \theta_m, \tau_1,\ldots, \tau_m)$, $\tau_j \in \mathbb{R}_+$ and $\theta_j \in \mathbb{R}_+$ are the knots and power-law slopes for $j = 1,\dots, m$, respectively, and $\tau_{m+1} = +\infty$.  Together, (\ref{eq:observed_model})-(\ref{eq:lambda.prior}) form a hierarchical model for the source intensities.

In our analyses, we consider two- and three-component piecewise-linear $\log(N)-\log(S)$ relationships. As in  \cite{udaltsova2014universe} and \cite{stampoulis2017bayesian}, we specify gamma prior distributions on the power-law slopes, i.e., $\theta_j \sim \operatorname{Gamma}(a_j,b_j)$, for $j=1,\ldots,m$. We specify gamma and truncated Normal prior distributions on the lower bound and breakpoints, with truncation imposed as needed to preserve their ordering, $\tau_j < \tau_{j+1}$. The specific prior distributions and hyperparameters used in our analyses are described in the Supplement.


\section{Poisson-binning 
for non-ignorable missing data}\label{section:poisson_binning}

We model survey incompleteness through a missing data mechanism at the population level of the hierarchical model, i.e., at the level of the source intensity distribution in (\ref{eq:lambda.prior}). Under the observed counts model (\ref{eq:observed_model}), the probability that any given source is detected is a function of its count rate, background rate, and survey-specific characteristics.

Let $n$ denote the observed number of detected sources. The full posterior distribution under the resulting hierarchical model can be written as
\begin{align}\label{form:full_posterior}
p(\pmb{\lambda}, \xi, \pmb{\theta}, \pmb{\phi}, \psi 
\mid X, \pmb{Y}, n)
\propto 
p(X, \pmb{Y} \mid \pmb{\lambda}, \xi) \;
p(\pmb{\lambda}, n \mid \pmb{\theta}, \pmb{\phi}, \psi) \;
p(\xi, \pmb{\theta}, \pmb{\phi}, \psi),
\end{align}
where $\pmb{\lambda} = (\lambda_1,\dots,\lambda_n)$ denotes the intensities of the detected sources,  $\pmb{\phi}$ parameterizes the missing data mechanism, and $\psi$ denotes the expected total number of sources in the survey region (see Section~\ref{section:binning_one_survey}). The factorization follows from the conditional independence structure of the hierarchical model. 
The first term on the right-hand side of (\ref{form:full_posterior}) corresponds to the object-level model in (\ref{eq:observed_model}) and (\ref{eq:background_model_uniform}), and the last term to prior distributions.

In this section, we focus on the middle term $
p(\pmb{\lambda}, n \mid \pmb{\theta}, \pmb{\phi}, \psi),$
which captures the effect of survey incompleteness. For this purpose, we condition on $\pmb{\lambda}$, or alternatively, treat them as observed data. This conditional model is built into the hierarchical model in Section~\ref{section:incompleteness}.

\subsection{A single incomplete survey}\label{section:binning_one_survey}

To account for an incomplete survey, suppose that $\lambda_i \indep p_\Lambda(\lambda_i \mid \pmb{\theta})$, for $i = 1, \dots, N$, where $N \sim \text{Pois}(\psi)$ but only a subsample of the complete set of sources are detected. Specifically, suppose $Z_i \mid \lambda_i, \pmb{\phi}  \sim \text{Bernoulli}(g(\lambda_i; \pmb{\phi}))$ and we observe source $i$ if and only if $Z_i=1$. Here, 
$n= \sum_i Z_i$ is the observed sample size, 
$\psi$, $\pmb{\theta}$, and $\pmb{\phi}$ are unknown parameters, the latter two of which may be vectors, and we refer to $g(\lambda;\pmb{\phi})$ as the incompleteness function. This missing data mechanism is non-ignorable because the incompleteness function depends on $\lambda$. Thus, simply treating the detected sources as a sample of $p_\Lambda(\lambda| \pmb{\theta})$ would be biased. Instead, the density for the observed sample is 
\begin{align}\label{form:observed_density}
    p_{\Lambda|Z = 1} (\lambda_i \mid \pmb{\theta},\pmb{\phi}) = \frac{g(\lambda_i; \pmb{\phi}) \ p_\Lambda(\lambda_i|\pmb{\theta}) }{\int g(u;\pmb{\phi}) \ p_\Lambda(u| \pmb{\theta})  {\rm d}u  }.
\end{align}
Because the denominator, the marginal detection probability  $\pi(\pmb{\theta}, \pmb{\phi}) = \int p_\Lambda(u| \pmb{\theta}) g(u; \pmb{\phi}) {\rm d}u$, depends on $(\pmb{\theta}, \pmb{\phi})$ it cannot be ignored, e.g., when computing the likelihood. In a typical iterative optimizer (e.g., for maximum likelihood) or posterior sampler, $\pi(\pmb{\theta}, \pmb{\phi})$ must be computed repeatedly, potentially several times per iteration. If $g$ is only available in a lookup table (as common in astronomy), this integration involves computing the integrand on a fine grid in $u$.  

By the thinning property of Poisson processes, $n \sim \text{Pois}(\psi \pi(\pmb{\theta}, \pmb{\phi}) )$ and  
combining the density of detected intensities in (\ref{form:observed_density}) with the Poisson likelihood for $n$, the $n\log \pi(\pmb{\theta},\pmb{\phi})$ terms cancel, yielding the joint log-likelihood 
\begin{align}\label{form:unbinned_likelihood}
    l(\psi,\pmb{\theta},\pmb{\phi}; \pmb{\lambda})  
    =&  \sum_{i=1}^n \log(g(\lambda_i; \pmb{\phi})) 
    + \sum_{i=1}^n \log(p_\Lambda(\lambda_i| \pmb{\theta})) 
    + n \log(\psi) - \psi \pi(\pmb{\theta}, \pmb{\phi}).
\end{align}
We refer to (\ref{form:unbinned_likelihood}) as the \textit{target} likelihood.

\subsubsection{Poisson-binning likelihood} 
To avoid repeated numerical evaluation of the integral 
$\pi(\pmb{\theta}, \pmb{\phi})$, we consider a discretized (binned) formulation of the model.  
Let $\mathcal{B}_1, \dots, \mathcal{B}_B,$ be a set of bins that partition the support of $\lambda$;\footnote{If the $\lambda$-support is unbounded, assume $\int_{\bigcup \mathcal{B}_k} p_\Lambda(\lambda| \pmb{\theta}) {\rm d}\lambda > 1 -\epsilon$ for some small $\epsilon$ and all  admissible $\pmb{\theta}$. 
} 
let  $b_k$ and $\delta_k$ be the midpoint and width of bin $\mathcal{B}_k$, respectively. We reduce the complete and observed data to bin counts, 
\begin{align}
X_k^{\text{com}} &= \#\{\lambda_i \in {\cal B}_k \}\\
X_k^{\phantom{com}} &= \#\{\lambda_i \in {\cal B}_k \hbox{ with } Z_i=1\},
\end{align}
and formulate a Poisson-binning model as
\begin{eqnarray}
    N &\sim& \text{Pois}(\psi) \label{eq:Ndist}\\
    (X_1^{\text{com}}, \dots , X_B^{\text{com}}) &\sim& \text{Multinomial}\Big(N, (p_\Lambda(b_1| \pmb{\theta}) \delta_1, \, \ldots, \, p_\Lambda(b_B| \pmb{\theta}) \delta_B) \Big) \label{form:binning_com_multinomial} \\
    X_k | X_k^{\text{com}} &\indep& \text{Binomial}\Big(X_k^{\text{com}}, g(b_k; \pmb{\phi})\Big). \label{form:binning_obs_binomial}
\end{eqnarray}
It follows directly that the observed bin counts are distributed
\begin{align}\label{form:X_k_Poisson}
     X_k \overset{\text{ind.}}{\sim} \text{Pois}\Big(\psi \, 
     g(b_k; \pmb{\phi}) \,
     p_\Lambda(b_k \mid \pmb{\theta}) \,  \delta_k\Big)
\end{align}
and are independent of the unobserved bin counts, $\{X_k^\text{com} - X_k, k=1,\ldots, B\}$. Thus, we can base inference on (\ref{form:X_k_Poisson}) via the Poisson-binned log-likelihood, 
\begin{align}\label{ref:likelihood_binned}
    l^*(\psi,\pmb{\theta}, \pmb{\phi}; \pmb{\lambda}) &= 
    \sum_{k=1}^B X_k \log(\psi \, g(b_k; \pmb{\phi}) \,
    p_\Lambda(b_k\mid \pmb{\theta}) \, \delta_k)
    - \psi \sum_{k=1}^B  g(b_k; \pmb{\phi}) \, 
    p_\Lambda(b_k\mid \pmb{\theta}) \, \delta_k  \nonumber\\
    & =  \sum_{k=1}^B X_k \log(g(b_k;\pmb{\phi})) 
        + \sum_{k=1}^B X_k \log( p_\Lambda(b_k\mid \pmb{\theta})) 
        +n \log(\psi) 
        - \psi\, \hat\pi(\pmb{\theta},\pmb{\phi})
        + c,
\end{align}
where $\hat\pi(\pmb{\theta},\pmb{\phi}) =  
\sum_{k=1}^B g(b_k; \pmb{\phi}) \, p_\Lambda(b_k\mid \pmb{\theta}) \,  \delta_k$.

\subsubsection{Efficient likelihood approximation}
\label{sec:approx}
The Poisson-binning likelihood in (\ref{ref:likelihood_binned}) and the target likelihood in (\ref{form:unbinned_likelihood}) are equal in the limit as the binning becomes arbitrarily fine. Suppose, for example, that we consider equal-width bins and let the number of bins grow. In the limit,  
\begin{align}
    &
    \sum_{k=1}^B  g(b_k;\pmb{\phi})\delta_k \, p_\Lambda(b_k\mid \pmb{\theta}) 
    =\hat\pi(\pmb{\theta}, \pmb{\phi})
    \rightarrow \pi(\pmb{\theta}, \pmb{\phi}),  \label{form:approx_cond_i} \\
    &
    \text{and } X_k \in \{0, 1\} \text{ for all } k, 
    \label{form:approx_cond_ii}
\end{align}
and thus,
\begin{align}
    \label{form:approximate_likelihoods} 
     l^*(\psi, \pmb{\theta}, \pmb{\phi}; \pmb{\lambda}) &\rightarrow 
       \sum_{i=1}^n \log(g(\lambda_i ;\pmb{\phi})) 
     +  \sum_{i=1}^n \log(p_\Lambda(\lambda_i \mid \pmb{\theta})) 
     + n \log(\psi) 
     - \psi \pi(\pmb{\theta}, \pmb{\phi}) 
     \\
   &    = l(\psi,\pmb{\theta}, \pmb{\phi};\pmb{\lambda}). \nonumber
\end{align}
In (\ref{form:approximate_likelihoods}), we evaluate $p_\Lambda$ and $g$ for the observed sources (at the unique $\lambda_i \in \mathcal{B}_k$)  
instead of the bin midpoints $b_k$. This is possible since there is at most one $\lambda_i$ in each bin, see (\ref{form:approx_cond_ii}).

In practice, one may choose a binning scheme that is sufficiently fine to ensure (\ref{form:approx_cond_ii}). Indeed, Poisson-binning can be viewed as a computationally efficient approximation to the target likelihood in (\ref{form:unbinned_likelihood}) even when (\ref{form:approx_cond_ii}) does not hold, for example, when some $X_k > 1$. In high-density regions,
$\sum_{i: \lambda_i \in \mathcal{B}_k} \log(p_\Lambda(\lambda_i| \pmb{\theta}))$ may be well approximated by $X_k \log( p_\Lambda(b_k| \pmb{\theta}))$ for a sufficiently narrow bin.
(Likewise, $\sum_{i: \lambda_i \in \mathcal{B}_k} \log(g(\lambda_i; \pmb{\phi}))$ may be well approximated by $X_k \log( g(b_k; \pmb{\phi}))$, particularly in regions where $g$ varies slowly.) In practice, we may choose a coarse binning in the early iterations of a likelihood optimizer or posterior sampler, and increase the number of bins as the algorithm reaches convergence / stationarity.  
If the observed data are themselves measured with appreciable error, a correspondingly coarse binning may also be appropriate. 
The bins need not be of equal width; narrower bins may be placed in regions where $p_\Lambda$ or $g$ exhibit substantial curvature, and wider bins elsewhere.

\subsection{Multiple incomplete surveys}\label{section:binning_multiple_surveys}

Consider the general case where we obtain data from $J$ surveys and $Z_{ij} = 1$ is equal to one if source $i$ is observed in survey $j$ and zero otherwise. Source $i$ may be observed in multiple surveys and  
each survey 
has its own incompleteness function which we specify via  
\begin{align}
    g_1(\lambda_i; \pmb{\phi}) &=  \Pr(Z_{i1} = 1 | \lambda_i, \pmb{\phi}), \nonumber\\
    g_2(\lambda_i; \pmb{\phi}) &=  \Pr( Z_{i1}= 0 \text{ and } Z_{i2} = 1 \mid \lambda_i, \pmb{\phi}), \nonumber\\
      & \ \vdots \label{eq:complete}\\
    g_J(\lambda_i; \pmb{\phi}) &=  \Pr\left( \sum_{j = 1}^{J-1} Z_{ij} = 0 \text{ and } Z_{iJ} = 1 \big| \lambda_i, \pmb{\phi} \right), \nonumber
\end{align}
where $\pmb{\phi}$ collects the parameters governing survey detection. 
We define the incompleteness functions so that their values are the probabilities of disjoint events, with each subsequent function recording the probability that a source is detected in a survey but does not appear in any of the previous surveys. Let ${\cal J}_i =\min\{j:Z_{ij}=1\}$  for sources 
detected in at least one survey 
(and be undefined for undetected sources). We set  $n_j = \sum_i \mathds{1}({\cal J}_i=j)$ to the number of sources observed in survey $j$, but not in surveys $1,\ldots, j-1$.

\subsubsection{Dependent incompleteness functions}\label{section:general_incompleteness}

Thus far, we have expressed the incompleteness functions, $g_j(\cdot)$, in terms of the source intensities $\lambda_i$. In practice, however, detection probabilities may depend on observed covariates, model parameters, or tuning parameters within the detection algorithms. Examples include the level of background contamination, the location on the detector where the source is observed, and the false positive rate of the detection algorithm. We discuss these variables in Section~\ref{section:incompleteness_marginalization}. Here, we generalize the model formulation to accommodate incompleteness functions that depend on them. 

Let $\pmb{W} =(\pmb{W}_1, \ldots \pmb{W}_n)$ be the observed covariates for each detected source. We assume that their distribution, $p_{\pmb{W}}(\pmb{w})$, (among the detected and undetected sources) is known.  In practice, this is reasonable since $\pmb{W}$ is composed of covariates that vary across detectors in a known fashion, and sources can be assumed to be uniformly distributed across the field of view.  We can also assume that $\pmb{W}$ and $\lambda$ are independent since $\pmb{w}$ describes characteristics of the detector and $\lambda$ is intrinsic to the astrophysical source.  We generalize the incompleteness functions given in (\ref{eq:complete}) to
\begin{align}
    \label{eq:general_complete}
    g_{j\pmb{w_i}}(\lambda_i; \pmb{\phi}, \xi) &=  \Pr\left( \sum_{\ell = 1}^{j-1} Z_{i\ell} = 0 \text{ and } Z_{ij} = 1 \big| \lambda_i, \pmb{\phi}, \xi, \pmb{w}_i\right),
\end{align}
for $j=1,\ldots, J$, 
where the parameter $\xi$ is separated from $\pmb{\phi}$ because it plays a specific role in the overall model, e.g., in the distribution of the background count given in (\ref{eq:background_model_uniform}).  

We aim to derive the density of $\lambda_i$ given ${\cal J}_i = j.$ By the law of iterated expectations, the probability that ${\cal J}_i=j$ is given by
$\pi_j(\pmb{\theta}, \pmb{\phi}, \xi) = \int  g_{j\pmb{w}}(\lambda; \pmb{\phi}, \xi) \, p_\Lambda(\lambda \mid \pmb{\theta}) \, p_{\pmb{W}}(\pmb{w}) \,{\rm d}\lambda\,{\rm d}\pmb{w}$ and thus the observed number of sources with ${\cal J}_i=j$ is $n_j\sim\text{Pois}(\psi \ \pi_j(\pmb{\theta}, \pmb{\phi}, \xi))$.  
Finally, the joint distribution of the detected source intensities and covariates given ${\cal J}=j$ is
\begin{align}\label{form:observed_density_cov}
    p_{\Lambda\pmb{W}|{\cal J}=j} (\lambda_i,\pmb{w}_i \mid \pmb{\theta},\pmb{\phi}, \xi) 
    = \frac{ g_{j\pmb{w}_i}(\lambda_i; \pmb{\phi},\xi) \, p_\Lambda(\lambda_i\mid\pmb{\theta}) \, p_{\pmb{W}}(\pmb{w}_i)}
    {\pi_j(\pmb{\theta}, \pmb{\phi}, \xi) };
\end{align}
combining this with the independent Poisson distributions of $n_1,\ldots n_J$ 
yields the target log-likelihood, up to an additive constant, given by 
\begin{align}\label{form:unbinned_likelihood_cov}
    l(\psi,\pmb{\theta},\pmb{\phi},\xi; \pmb{\lambda}) 
    =&  \sum_{i=1}^n \log(g_{j_i\pmb{w}_i}(\lambda_i; \pmb{\phi},\xi))
    + \sum_{i=1}^n \log(p_\Lambda(\lambda_i| \pmb{\theta})) 
    + n \log(\psi)
    - \psi \sum_{j=1}^J \pi_j(\pmb{\theta}, \pmb{\phi}, \xi), 
\end{align} 
where $j_i$ is the observed value of ${\cal J}_i$.
This likelihood makes clear the interaction between the population model and the survey detection mechanisms. In general, joint estimation of the expected source count $\psi$, the low-intensity tail of $p_\Lambda$, and flexible incompleteness functions $g_j$ may be only weakly identifiable; 
in our application, identifiability is strengthened by an externally calibrated X-ray incompleteness (detection) function.

\subsubsection{The full Poisson-binning likelihood} 
In practice, covariates may vary across the detector in ways that can be represented by a moderate number of bins. 
In fact, incompleteness functions are typically tabulated in astronomy, which effectively bins covariates \citep[e.g.,][]{wright2023simulating}. Let ${\cal C}_1, \ldots, {\cal C}_L$ be a set of bins that partitions the support of $\pmb{W}$ with $\pmb{c}_\ell$ and $\delta_\ell$ being the centroid and volume of bin ${\cal C}_\ell$, respectively.

The full Poisson-binning model can be formulated analogously to (\ref{eq:Ndist}) and (\ref{form:binning_com_multinomial}), but with 
\begin{align}  \label{eq:bincounts} 
X_{jk\ell} &= \#\{\lambda_i \in {\cal B}_k \hbox{ with } \pmb{W}_i\in {\cal C}_\ell, \ Z_{ij}=1, \text{ and } Z_{im}=0 \text{ for } m < j \},
\end{align}
$X_{+k\ell} = \sum_j X_{jk\ell}$,  and $X_{k\ell}^{\text{com}} = \#\{\lambda_i \in {\cal B}_k \hbox{ with } \pmb{W}_i\in {\cal C}_\ell \}$ modeled as
\begin{align} 
     X_{+k\ell} | X_{k\ell}^{\text{com}} &\sim \text{Binomial}\left(X_{k\ell}^{\text{com}}, g_{\pmb{c}_\ell}(b_k; \pmb{\phi}, \xi ) \right), \label{form:binning_obs_binomial} 
     \ \ \hbox{ and} \\
    (X_{1k\ell} , \dots , X_{Jk\ell}) \big|  X_{+k\ell} &\sim \text{Multinomial}\left(    X_{+k\ell} , 
    \left(
    \frac{g_{1\pmb{c}_\ell}(b_k; \pmb{\phi},\xi)}{g_{\pmb{c}_\ell}(b_k; \pmb{\phi},\xi )}, \dots, 
    \frac{g_{J\pmb{c}_\ell}(b_k; \pmb{\phi},\xi)}{g_{\pmb{c}_\ell}(b_k; \pmb{\phi},\xi) } \right) \right). \label{form:binning_obs_multinomial} 
\end{align}
It follows directly that 
\begin{align}\label{form:X_k_Poisson_multiple}
     X_{jk\ell} \indep \text{Pois}\Big(\psi \,  \, g_{j\pmb{c}_\ell}(b_k; \pmb{\phi}, \xi) \, p_\Lambda(b_k| \pmb{\theta}) \,p_{\pmb{W}}(\pmb{c}_\ell)\, \delta_k \,\delta_\ell\Big). 
\end{align}
Setting $\hat\pi_j(\pmb{\theta},\pmb{\phi},\xi) =
\sum_{k\ell}  g_{j \pmb{c}_\ell}(b_k; \pmb{\phi},\xi) 
           \, p_\Lambda(b_k| \pmb{\theta})
           \, p_{\pmb{W}}(\pmb{c}_\ell) \, \delta_k \, \delta_\ell$,
the Poisson-binned log-likelihood is, up to an additive constant, given by 
\begin{align}\label{ref:likelihood_binned_multiple}
    l^*(\psi,\pmb{\theta}, \pmb{\phi},\xi; \pmb{\lambda}) 
    =  \sum_{jk\ell} X_{jk\ell} 
    \log(\psi \, g_{j\pmb{c}_\ell}(b_k;\pmb{\phi},\xi) \,
    p_\Lambda(b_k| \pmb{\theta}) )  
    - \psi \,\sum_{j=1}^J\hat\pi_j(\pmb{\theta},\pmb{\phi},\xi), 
\end{align}
which can be viewed as an approximation of $l(\psi,\pmb{\theta}, \pmb{\phi},\xi)$, just as described in Section~\ref{sec:approx}.

\section{A hierarchical model for two incomplete surveys}\label{section:incompleteness}

Consider the scenario described in Section~\ref{section:Challenges_and_prospects} where we analyze a sample of sources, each of which is detected in either an X-ray or optical catalog, with both catalogs incomplete. In practice, the set of sources detected in the two surveys overlaps, and both surveys contain sources not detected in the other. Undoubtedly, there are also sources that are not detected in either survey. The model described in Section~\ref{section:source_model} can be used for a single complete survey and the model described in Section~\ref{section:poisson_binning} can be used if the source intensities $\lambda$ are known. Here, we combine these models into a hierarchical model that is appropriate for two incomplete catalogs, in which source intensities are inferred from observed background-contaminated photon counts.  
The aim is 
joint estimation of the source intensities, $\pmb{\lambda}$, and the distribution of the source population by rigorously accounting for incompleteness in the (X-ray and optical) surveys.

Let $n$ denote the number of sources detected in at least one of the two surveys. For each such source $i=1,\ldots,n,$ we extract the X-ray photon count $Y_i$ from the corresponding PSF-defined X-ray  region and record the associated covariate vector $W_i$. 
We also observe the X-ray background count, $X$, and the missing data indicator variables 
\begin{align} 
Z_{i1} &= 1 \text{ if and only if source } i  \text{ is detected in the X-ray catalog}  \nonumber\\
Z_{i2} &= 1 \text{ if and only if source } i  \text{ is detected in the optical catalog}.      \nonumber  
\end{align}
As defined in Section~\ref{section:binning_multiple_surveys}, we let ${\cal J}_i = \min\{j : Z_{ij}=1\}$ denote the survey index associated with source $i$. 
The areas of the X-ray source regions, $a_i$, their normalized effective areas, $e_i$, the exposure time, ${\cal T}$, and the area of the X-ray background region are also recorded. The model is parameterized in terms of the X-ray source intensities, $\lambda_i$, the background count rate, $\xi$, the parameters of the Pareto mixture, $\pmb{\theta}$, the parameters of the optical incompleteness function, $\pmb{\phi}$, and the expected number of sources $\psi$.  Accounting for the covariates, the full posterior distribution in (\ref{form:full_posterior}) generalizes to 
\begin{align}\label{form:full_posterior_with_cov}
p(\pmb{\lambda}, \xi, \pmb{\theta}, \pmb{\phi}, \psi 
\mid \pmb{W}, X, \pmb{Y}, \pmb{\cal J}) 
\propto 
p(X, \pmb{Y} \mid \pmb{\lambda}, \xi) \; 
p(\pmb{\lambda}, \pmb{W}, \pmb{\cal J} \mid \pmb{\theta}, \pmb{\phi}, \psi) \; 
p(\xi, \pmb{\theta}, \pmb{\phi}, \psi), 
\end{align}
where the terms on the right hand side are the object-level model, the population model (including the survey detection mechanism), and the joint prior distribution, respectively. 

We assume the object-level Poisson model in (\ref{eq:observed_model}) for the individual X-ray source counts and (independently) the Poisson model in (\ref{eq:background_model_uniform}) for the background X-ray count. These models are specified conditional on $\lambda$ and $\xi$. At the population level, we specify an incompleteness corrected Pareto mixture model for $\lambda$. The likelihood can either be formulated directly as in (\ref{form:observed_density_cov}) and (\ref{form:unbinned_likelihood_cov}) or via the Poisson-binned likelihood in
(\ref{eq:bincounts})--(\ref{ref:likelihood_binned_multiple}). 

In either case, we treat the X-ray detection function $g_{1\pmb{w}}(\lambda)$ as known. It is computed in advance from a comprehensive simulation study of the X-ray survey detection pipeline (see \citealp{wright2023simulating}) and 
has no unknown parameters. The optical detection function is modeled parametrically as $g_{2\pmb{w}}(\lambda;\pmb{\phi}_2)$ and its parameters $\pmb{\phi}_2$ are estimated,
see Section~\ref{section:incompleteness_marginalization}; throughout our analysis, $\pmb{\phi} \equiv \pmb{\phi}_2$.
We assume $\pmb{\phi}, \pmb{\theta}, \xi,$ and $\psi$ to be a priori independent. The specific prior distributions used in Sections~\ref{section:sim_studyLogNlogS} and~\ref{section:CDFS_posterior} are described in the Supplement.

\subsection{The incompleteness functions}\label{section:incompleteness_marginalization}

Here we specify the survey-specific incompleteness functions used in the applications in Section~\ref{section:CDFS_analysis}. 
As discussed in Section~\ref{section:general_incompleteness}, incompleteness functions typically depend on observed covariates. 
In our numerical studies, for example, we use the X-ray incompleteness function of \cite{wright2023simulating}, which is based on extensive simulations designed to mimic Chandra X-ray survey data. This incompleteness function depends on the expected background count per pixel $\xi_T = \mathcal{T}\xi$, on the off-axis angle, $L_i$, which describes the distance from the source location to the center of the field of view, on the effective area $e_i$, and on $w_{0}$, a threshold that determines the sensitivity of the X-ray source detection algorithm wavdetect \citep{freeman2002wavelet}; see Section~\ref{section:CDFS_data}. We treat $(L_i, e_i)$ as observed covariates, $\mathcal{T}$ and $w_0$ as constants, 
and $\xi$ as a model parameter. Specifically, the X-ray incompleteness function 
of \cite{wright2023simulating} can be expressed as
\begin{align}\label{form:g_fct_wright_lambda} 
 g_{1\pmb{w}}(\lambda) =    
 \Pr(Z_{i1} = 1 \mid \lambda_i, \xi, L_i, e_i) = 
 1 - \exp \left( \frac{-(\lambda_i \mathcal{T} e_i)^{\nu_1(\xi \mathcal{T} , \, L_i)}}{10^{\nu_2(\xi \mathcal{T} , \,L_i)}}  \right), 
\end{align}
where  $\nu_1( \xi_T, \, L)$  and $\nu_2( \xi_T,\, L)$ are given by \cite{wright2023simulating} in tabular form for various combinations of $\xi$ and $L_i$.\footnote{This formulation in (\ref{form:g_fct_wright_lambda}) is a 
notational adaptation of the detection function given by  
\cite{wright2023simulating}. For example, we suppress the dependence of $\nu_1$ and $\nu_2$ on $\mathcal{T}$ and $w_0$ which are constants in our setting.  
\cite{wright2023simulating} formulate the incompleteness function in terms of the expected source counts, $C$, rather than $\lambda$ (count/s). To obtain the formulation in (\ref{form:g_fct_wright_lambda}), we set $C_i = \lambda_i \mathcal{T} e_i$. We also use $\nu_1(\cdot)$ and $\nu_2(\cdot)$ instead of $\lambda_1(\cdot)$ and $\lambda_2(\cdot)$ of \cite{wright2023simulating} to avoid a notational clash with source intensities.} 
We use bilinear interpolation over the calibration grid in $(\xi_T,L)$, at the level of the detection probability to evaluate $g_{1\pmb{w}}(\lambda)$. 
Since the background region contains a large number of counts, $\xi$ is tightly constrained. For evaluation of $g_{1\pmb{w}}$, we use the empirical estimate $\hat\xi = X/(A\mathcal{T})$, while retaining $\xi$ as a parameter in the photon-count model. The resulting approximation has negligible impact on the incompleteness correction.

We leverage the carefully calibrated X-ray incompleteness function of \cite{wright2023simulating} to inform estimation of the optical incompleteness function, $g_{2\pmb{w}}(\lambda ; \pmb{\phi}_2).$ (Since $g_{1\pmb{w}}(\lambda)$ has no unknown parameters $\pmb{\phi}= \pmb{\phi}_2$.)  
The number of sources detected only in the optical survey in each source intensity bin given the total number of sources detected in that bin is given in (\ref{form:binning_obs_multinomial}) and simplifies in the case of two surveys to
\begin{align}
    X_{2k\ell} \big|  X_{+k\ell}
    \sim \text{Binomial}\left(
    X_{+k\ell},
    \frac{g_{2\pmb{c}_\ell}(b_k; \pmb{\phi}_2)}
         {g_{1\pmb{c}_\ell}(b_k) + g_{2\pmb{c}_\ell}(b_k; \pmb{\phi}_2 )}
    \right).
\end{align}
Thus, $X_{2k\ell}/(X_{1k\ell}+X_{2k\ell})$ can be viewed as an estimator of 
$
\frac{g_{2\pmb{c}_\ell}(b_k; \pmb{\phi}_2)}
     {g_{1\pmb{c}_\ell}(b_k) + g_{2\pmb{c}_\ell}(b_k; \pmb{\phi}_2 )},
$
which in turn can be used to constrain $g_{2\pmb{c}_\ell}(\lambda; \pmb{\phi}_2)$ since $g_{1\pmb{c}_\ell}(b_k)$ is fully specified. In our formulation, $g_{2\pmb{w}}$ depends  on the covariates only through $g_{1\pmb{w}}$, and we use the parameterized model
\begin{equation}
g_{2\pmb{w}}(\lambda; \pmb{\phi}_2) = \big(1-g_{1\pmb{w}}(\lambda)\big) 
\left(1-\exp\!\left[-\left(\frac{\lambda}{\phi_{22}}\right)^{\phi_{21}}\right]\right),
\label{eq:g2-model}
\end{equation}
where $\pmb{\phi}_2 =(\phi_{21}, \phi_{22})$.  The first factor in (\ref{eq:g2-model}) reflects the survey ordering, accounting for the fact that $g_{2\pmb{w}}$ represents optical detections among sources \textit{not detected in the X-ray survey}.  
The second factor is motivated by the functional form of (\ref{form:g_fct_wright_lambda}). It has a Weibull cumulative distribution function form, representing the increase of detection probability with source intensity. The parameter $\phi_{21}$ controls the steepness of the transition from low to high probability, while $\phi_{22}$ sets the characteristic  scale at which optical detection becomes likely.

\subsection{Efficient implementation via Poisson-binning} \label{sec:implement}
The formulations of (\ref{form:unbinned_likelihood_cov}) and (\ref{ref:likelihood_binned_multiple}) depend on the marginal distribution $p_{\pmb{W}}(\pmb{w})$. 
We approximate $p_{\pmb{W}}$ using a Monte Carlo sample obtained from locations drawn uniformly from the exposure map of the analyzed CDF-S region. The resulting empirical distribution of $(L,e)$ provides a numerical approximation to $p_{\pmb{W}}$ under the assumption that sources are uniformly distributed over the field of view.
In this sense, the Monte Carlo sampling of $\pmb W$ induces a fine covariate binning, with each sampled $(L,e)$ serving as its own bin for evaluation and averaging of detection probabilities.

Beyond its statistical interpretation, Poisson-binning simplifies computation. 
By restricting all $\lambda$-dependence to the fixed grid $\{b_k\}$, the binned likelihood requires evaluations of the incompleteness functions only at these grid points. This allows all source-specific X-ray detection probabilities to be precomputed and stored as a lookup table prior to posterior sampling. 
Likewise, this discretization allows marginalization over the covariates $\pmb W$ to be reduced to bin-wise Monte Carlo averages that can be precomputed.
Each MCMC iteration only requires evaluation of the population model and the optical detection component on the grid, followed by a sum over bins. Repeated interpolation and Monte Carlo averaging inside the sampler are thus avoided, substantially reducing the computational cost of posterior sampling.  
Further details are provided in the Supplement.


\section{CDF-S data analysis}\label{section:CDFS_analysis}

We apply \name\ to obtain an incompleteness-corrected \lns\ fit using 
X-ray data from the Chandra Deep Field South (CDF-S), augmented with 
optical source locations from \cite{arnouts2001eso}. Throughout the 
analysis we compare an X-ray-only analysis, based only on the X-ray-detected sources, with the full \name\ model that incorporates both X-ray and optical information to assess the impact of the optical sources on posterior inference.

Section~\ref{section:CDFS_data} describes the data 
acquisition and preprocessing. 
In Section~\ref{section:sim_studyLogNlogS}, we conduct comprehensive 
and realistic simulation studies to evaluate the performance of the 
method. Finally, in Section~\ref{section:CDFS_posterior}, we apply \name\ to 
the CDF-S data and present posterior inference and \lns\ fits under 
the double and triple Pareto models.

\subsection{Chandra Deep Field South (CDF-S)}\label{section:CDFS_data}
CDF-S was obtained with the Chandra X-ray Observatory \citep{weisskopf2002overview}, a NASA-operated space telescope with very high sensitivity to faint X-ray sources. To obtain CDF-S, \chandra\ was pointed at the same patch of sky for a total of $\sim 44$ days  over several observation periods. We analyze a dataset that merges CDF-S observations from 2000, 2007 and 2010\footnote{ 
Merged CDF-S data are available at
\href{https://cxc.harvard.edu/cda/Contrib/CDFS.html}{link}. Its Data Identifier is: $\text{ADS/Sa.CXO\#Contrib/ChandraDeepFieldS}$. 
}
with a combined exposure time of $\mathcal{T} = 3.8\times10^{6}$ seconds. Figure~\ref{figure:CDFS_event_file}  
illustrates the CDF-S event file with X-ray counts plotted in white.  We consider a subregion, $\mathcal{S}$, of the dataset that has a total area of 1,168,052 pixels;
$\mathcal{S}$  is plotted as a red circle in Figure~\ref{figure:CDFS_event_file}.\footnote{We consider a circular area of radius 5 arcmin around the center of the CDF-S observations, given at RA= $03:32:28.2$ and DEC = $-27:48:36.0$ (FK5).}

We use the popular wavdetect \citep{freeman2002wavelet} X-ray source detection algorithm to identify sources in the CDF-S X-ray data. Using a significance 
threshold of $w_{0}= 10^{-6}$ to control the expected number of spurious detections,
we find a total of 189 X-ray sources in subregion $\mathcal{S}$.\footnote{
Source detection is performed on the $0.5$--$7$ keV band using wavelet scales 
$2,4,8,16,$ and $32$ pixels and the corresponding exposure map to account for spatial variations in exposure across the field. A threshold of $10^{-6}$ corresponds to roughly one spurious detection per $10^{6}$ pixels. 
The PSF map is constructed assuming an effective energy of $1.5$ keV 
and encircled counts fraction $0.393$, following CIAO procedures \citep{fruscione2006ciao}.} 
The associated regions obtained using the appropriate \chandra\ PSFs are illustrated as yellow circles for sources in a small magnified patch of ${\cal S}$  
in Figure~\ref{figure:patch_of_CDFS_all_sources}.

We obtain an optical catalog for \subregion\ using
the source list of \cite{arnouts2001eso}, which lists 45,912 detected optical (v-band) sources\footnote{The full optical data catalog is available here:  \href{https://vizier.cds.unistra.fr/viz-bin/VizieR?-source=J/A+A/379/740}{link}}. Of these, 2814 sources are located in \subregion.  
Restricting to the optically complete sample ($v<24.5$; \citealt{arnouts2001eso}) yields 1,316 sources in \subregion.
Note that completeness in the optical catalog does not imply completeness in the X-ray data considered here, since optical brightness and X-ray emission are not in one-to-one correspondence. 
The sky coordinates of the optical sources are used to construct the optical source regions used in our analysis. Specifically, circular regions are defined using the X-ray PSF, with radius set to three times the local PSF size (corresponding to an encircled counts fraction of approximately $r_i=0.99$).

The source regions are used to cross-match X-ray and optical sources. 
An optical source is matched to an X-ray source if its angular separation from the X-ray position is within the corresponding source-region radius.  
Overlapping source regions complicate the statistical model because the observed photon counts must then be partitioned among multiple sources. We therefore retain only sources with unique optical counterparts and non-overlapping source regions. A more general treatment of overlapping sources is the subject of ongoing work.

We further restrict analysis to X-ray sources with optical counterparts. 
The CDF-S data contain a mixture of source populations (e.g., active galactic nuclei, star-forming galaxies, and Galactic stars), and requiring an optical counterpart reduces contamination, yielding a more homogeneous sample for inference. This restriction is dataset specific and not required by the statistical model, which can accommodate X-ray sources without optical detections. 
The final sample contains $n=692$ optical sources within $\mathcal{S}$, of which 51 are detected in X-ray and optical data ($ \mathcal{J}_i =1$) and 641 only in the optical survey ($\mathcal{J}_i=2$).

For each retained source $i$, we extract the number of X-ray photons $Y_i$ and the source area $a_i$ (in pixels) from its source region in the CDF-S event file. 
Normalized effective areas $e_i$ are computed from the exposure map by averaging the exposure within each source region and normalizing to the aim point exposure. The exposure map and retained sources are illustrated in Figure~\ref{figure:exposure_map_main}. 
Background counts are estimated from the remaining area of $\mathcal{S}$ after removing all candidate source regions, yielding $X = 564,360$ counts over an area $A = 905,544$ pixels.

\subsection{Simulation study}\label{section:sim_studyLogNlogS}

To validate \name, we simulate 
data  under the full hierarchical model described in 
Section~\ref{section:incompleteness}, using the double-Pareto population model in \eqref{eq:lambda.prior}, the photon-count model \eqref{eq:observed_model}, and the survey-specific incompleteness functions \eqref{form:g_fct_wright_lambda} and \eqref{eq:g2-model}. The resulting simulation design closely reflects the geometry, background structure, and detection characteristics of the CDF-S survey.

We generate the global parameters $(\pmb\theta,\psi,\xi,\pmb\phi_2)$ 
from their joint posterior distribution obtained from the double-Pareto CDF-S analysis.  
We sample parameters from their posterior rather than the diffuse prior because prior draws are frequently unrealistic or unphysical, e.g., background rates outside the calibration range of the incompleteness model. Posterior-based parameters ensure realistic survey settings while retaining variability across simulation runs.

For each simulation replicate, the total number of sources is drawn from $N \sim \text{Poisson}(\psi)$. Source intensities $\lambda_i$ are then generated independently from the double-Pareto population model \eqref{eq:lambda.prior}. To reflect the empirical characteristics of the CDF-S data, we force the simulated populations to include two sources with intensity exceeding the upper break parameter $\tau_2$, thereby ensuring representation of the bright-end regime. Survey covariates $(L_i,e_i,a_i)$ 
are sampled with replacement from the CDF-S Monte Carlo sample described in Section~\ref{sec:implement}, with $a_i$ determined by the local PSF. 
This preserves the survey geometry and exposure characteristics of the CDF-S data. Observation time $\mathcal{T}$ and background area $A$ are fixed at the values obtained from the CDF-S data. Background counts are generated as $X \sim \text{Poisson}(A\,\mathcal{T}\,\xi)$.

Conditional on $(\lambda_i,\xi,L_i,e_i,\pmb{\phi}_2)$ for $i=1,\ldots,N$, survey detection indicators are generated using the X-ray incompleteness function \eqref{form:g_fct_wright_lambda} and the optical model \eqref{eq:g2-model}, thereby implementing the sequential survey sampling mechanism. Only sources detected in at least one survey are retained. 
For detected sources, photon counts are generated under  the Poisson model \eqref{eq:observed_model}.


\begin{figure}[t]
\centering
\includegraphics[width=0.49\linewidth]
{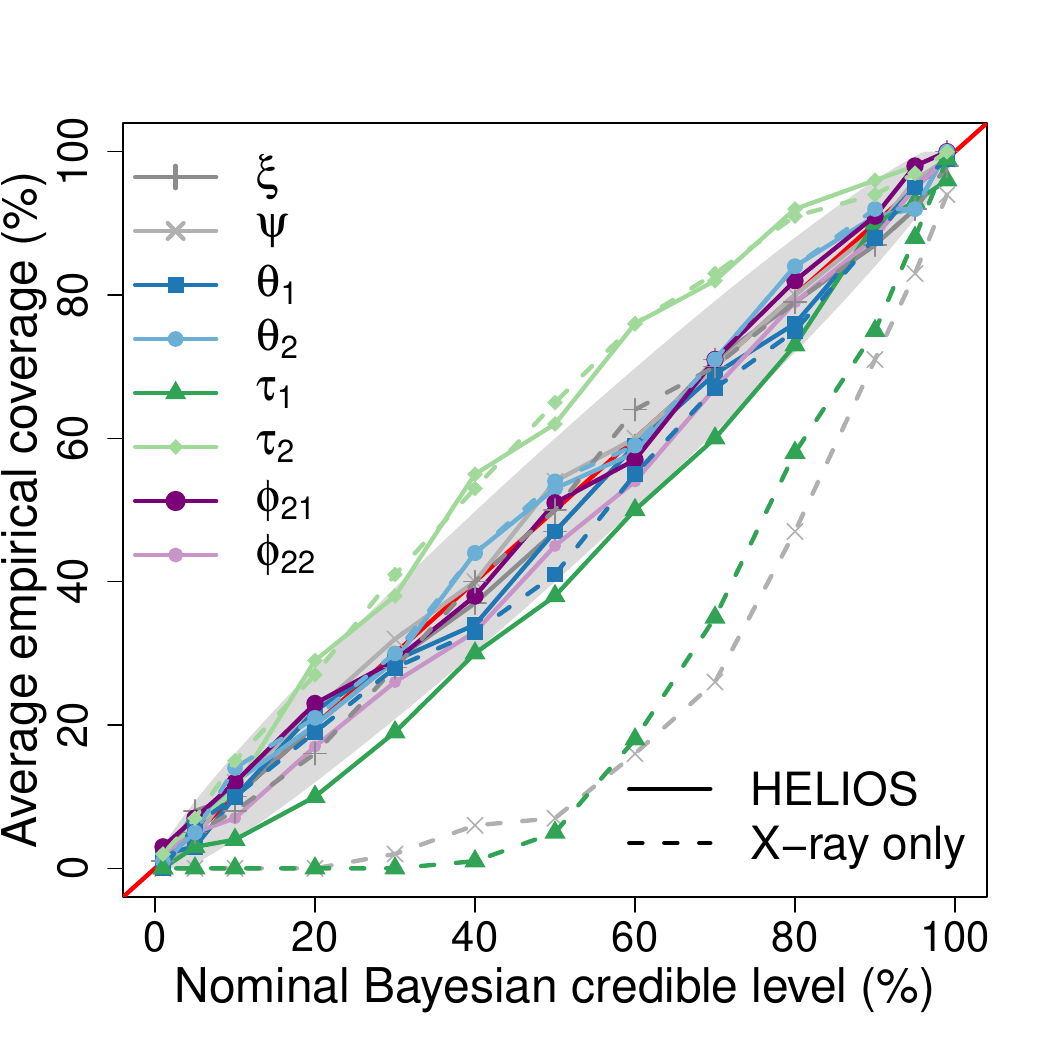}
\hfill
\includegraphics[width=0.49\linewidth]{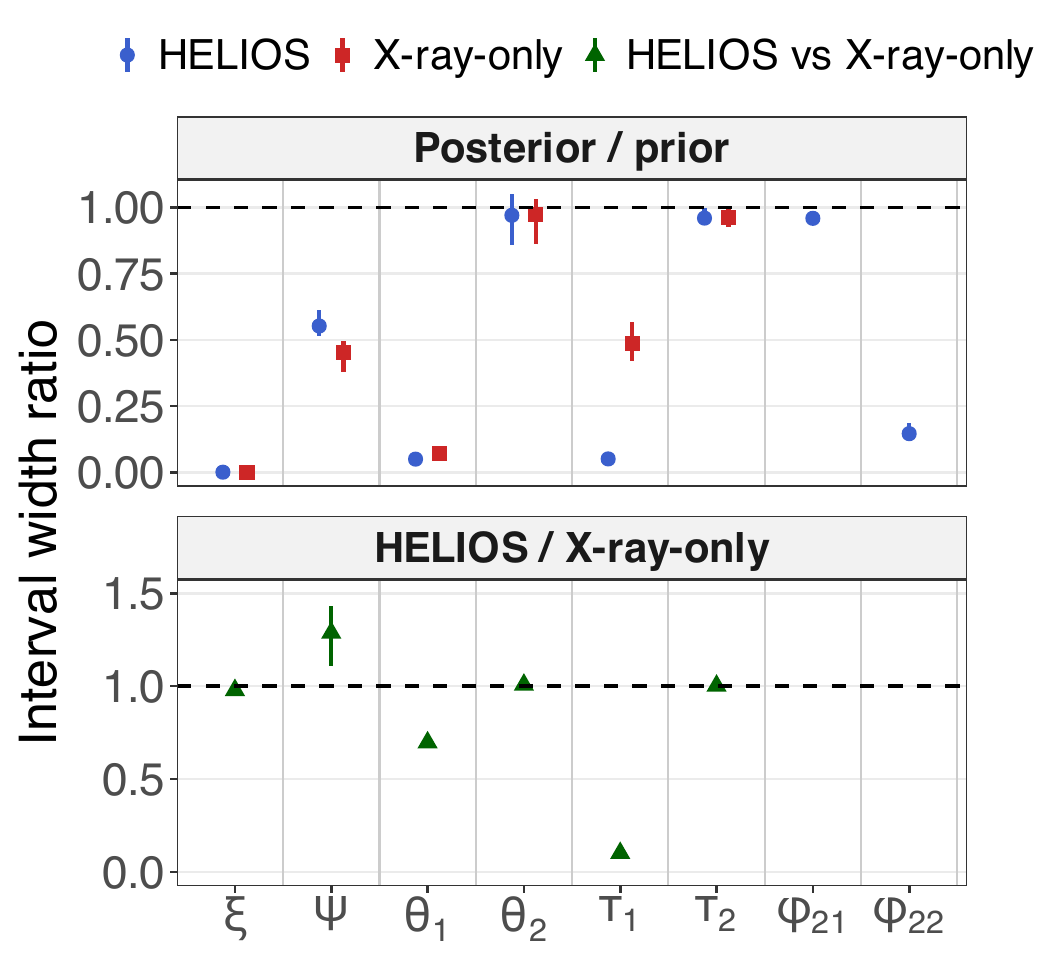} 
\caption{
Left: Empirical coverage of posterior credible intervals across simulation replicates against nominal credible level. Solid lines correspond to the full \name\ model (X-ray and optical data) and dashed lines to the X-ray-only model.
The red diagonal indicates ideal calibration, and the grey band the $95\%$ Monte Carlo uncertainty region under perfect calibration. 
Right: Posterior information content across models. The upper panel shows the ratio of posterior to prior $95\%$ interval widths for each parameter, 
measuring posterior contraction relative to the prior. 
The lower panel shows the ratio of posterior interval widths under the full and X-ray-only models. Values below one indicate narrower credible intervals under \name. Median ratios across simulation replicates and their interquartile
ranges are plotted. 
}
\label{fig:coverage_simulations}
\end{figure}

We generate 100 independent data sets and 
analyze each under an X-ray-only model using only the X-ray detected sources and the full \name\ model, which jointly incorporates both X-ray-detected and optical-only sources.  
This repeated-sampling design enables assessment of empirical credible-interval coverage and posterior precision under realistic survey conditions.  
Each simulated data set is fit using the Metropolis-within-Gibbs sampler detailed in the Supplement. Posterior samples are obtained from a single chain of $10^6$ iterations with the first $2\times10^5$ discarded as burn-in.

Figure~\ref{fig:coverage_simulations} shows the empirical coverage of the posterior credible intervals across simulation replicates.
For \name, empirical coverage closely follows the nominal credible levels for all parameters, indicating well-calibrated posterior uncertainty even with substantial survey incompleteness. In contrast, the X-ray-only model exhibits systematic undercoverage for the parameters $\tau_1$ and $\psi$. 
With only X-ray detected sources, the data contain relatively little information about very faint sources, which tends to push the estimated lower bound $\tau_1$ toward larger values.
A larger $\tau_1$ truncates more of the faint end of the population distribution, so a smaller population size $\psi$ is required to explain the observed detections. This leads to undercoverage of the credible intervals for both parameters.  
Incorporating optical information provides additional constraints on the faint-end regime, where X-ray detections are sparse, improving the calibration of the posterior distributions. 
Importantly, this undercoverage does not propagate to the slope parameter of primary interest $\theta_1$, whose credible intervals are well calibrated even under the X-ray-only model.

The upper right panel of Figure~\ref{fig:coverage_simulations} compares the width of the prior and posterior $95\%$ intervals;  
values below one indicate learning from the data. 
Both models produce substantial posterior contraction for 
$\xi$, $\theta_1$, and $\tau_1$,
and moderate contraction for $\psi$. 
The full \name\ model also shows substantial posterior contraction for the characteristic incompleteness scale  $\phi_{22}$. 
In contrast, the posterior intervals of the bright-end parameters $\theta_2$ and $\tau_2$ are similar to their prior intervals, reflecting the few bright sources constraining this part of the model.  
Posterior correlations between bright- and faint-end parameters are weak, suggesting that limited identification of the bright-end component has little or no effect on faint-end inference.

The lower right panel of Figure~\ref{fig:coverage_simulations} compares posterior interval widths under \name\ and the X-ray-only model. Incorporating optical information substantially improves precision for parameters governing the faint-end regime. In particular, the posterior interval for the key parameter $\theta_1$ is approximately $70\%$ as wide under \name\ as under the X-ray-only model, indicating a substantial gain in precision.
An even stronger improvement is observed for the breakpoint parameter $\tau_1$.  
For the population size parameter $\psi$, the full model yields slightly wider credible intervals. The narrower X-ray-only intervals reflect an overconfident posterior that tends to be biased toward smaller values of $\psi$, leading to undercoverage. Incorporating the optical information yields better calibrated uncertainty.  
For bright-end parameters, 
including $\theta_2$ and $\tau_2$, posterior interval widths are largely unchanged across the two models, since these parameters are primarily constrained by the brightest X-ray sources. 
Overall, these results demonstrate that incorporating the optical survey improves inference for the faint end of the population model while preserving well-calibrated posterior uncertainty.  

\subsection{CDF-S posterior inference and \lnsn}\label{section:CDFS_posterior}

We apply the proposed model to the CDF-S data described in Section~\ref{section:CDFS_data}. We compare an X-ray-only analysis, based on the $n=51$ sources detected in the X-ray data, with the full \name\ analysis, which augments these sources with the optical-only sample, yielding a combined sample of $n=692$ objects, and accounts jointly for X-ray and optical incompleteness. 
Posterior inference is obtained using eight parallel chains of the Metropolis-within-Gibbs sampler described in the Supplement.  
Each chain is run for $10^6$ iterations, discarding the first $2\times10^5$ iterations as burn-in. 
For the Poisson-binning likelihood, we use $B=2000$ bins equally spaced on the $\log_{10}(\lambda)$ scale, providing finer resolution at the faint end where source density is highest and the incompleteness function changes most rapidly. 
Convergence diagnostics and trace plots are provided in the Supplement and support convergence of the chains. 

\begin{figure}
\begin{minipage}{\linewidth}
  \centering
  \hspace{0.3cm}
  \includegraphics[width=0.99\linewidth]{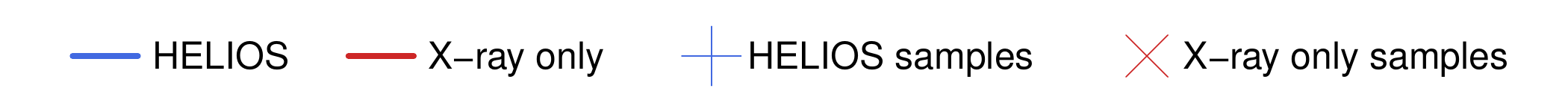}
  \includegraphics[width=0.51\linewidth]{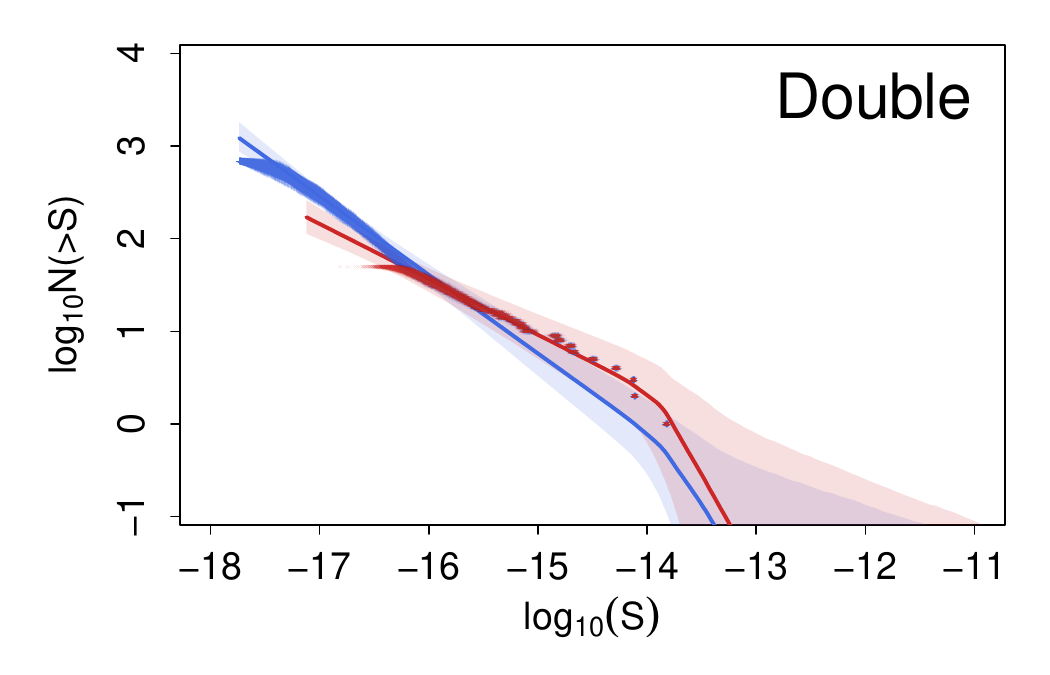}
  \hspace{-.7cm}
    \includegraphics[width=0.51\linewidth]{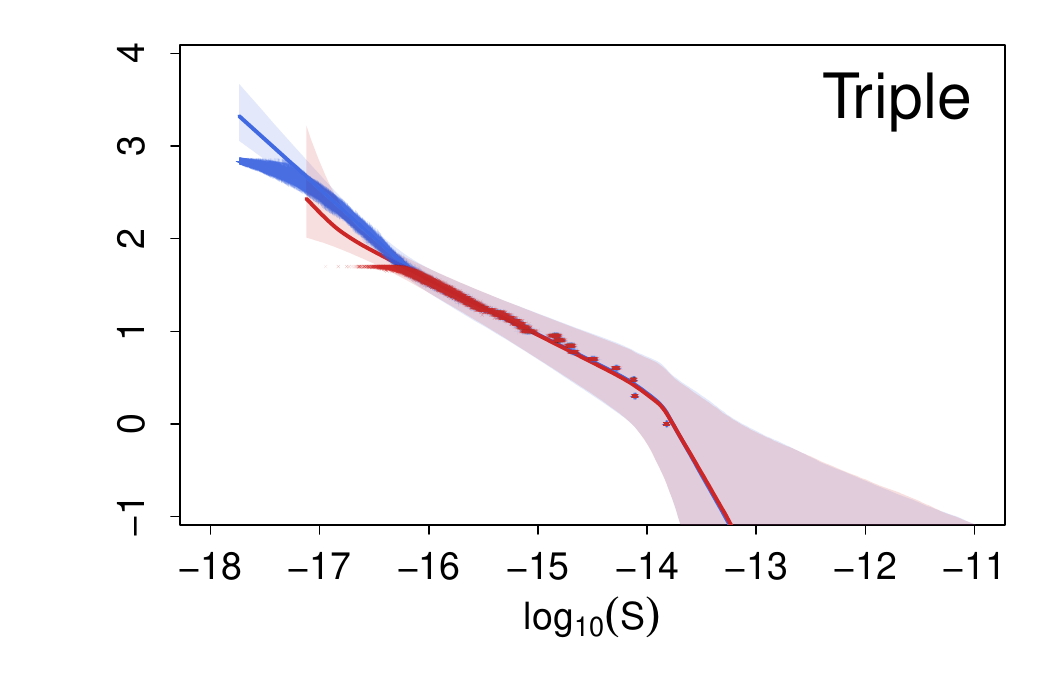}
    \caption{ 
Posterior \lns relations for the CDF-S data under the double Pareto model (left) and triple Pareto model (right). Solid curves show the posterior median of $\log_{10}N(>S)$ over the displayed threshold values $S$ for the full \name\ and X-ray-only fits, with shaded regions indicating pointwise $95\%$ credible intervals. Crosses show empirical \lns relations computed from posterior source-flux samples under each fitted model, treating each sampled set of source fluxes as complete.
}
    \label{figure:LogN-LogS}
  \end{minipage}
\end{figure}

Figure~\ref{figure:LogN-LogS} shows the posterior \lns relations implied by the fitted models. We report the inferred relations on the flux scale using the constant count-to-flux conversion factor $\gamma = 1.22\times10^{-11}\,\mathrm{erg\,cm^{-2}\,count^{-1}}$, so that $S_i=\gamma\lambda_i$ and $\tau_{Sj}=\gamma\tau_j$. Following standard practice for broken-power-law \lns estimates, we construct the curves conditional on a fixed lower flux bound, namely the marginal posterior median of $\tau_{S1}$ for the double Pareto model and of $\tau_{S0}$ for the triple Pareto model. We then obtain posterior draws of the remaining parameters conditional on this fixed value and compute $N(>S)=\psi\Pr(\Lambda_S>S\mid\boldsymbol{\theta})$ on a fixed grid of threshold values $S$, where $\Lambda_S=\gamma\Lambda$. Figure~\ref{figure:LogN-LogS} reports the pointwise posterior median of $\log_{10}N(>S)$ for the full \name\ and X-ray-only fits, with shaded regions giving the corresponding pointwise $95\%$ credible intervals. 
The crosses show empirical \lns relations computed from posterior source-flux samples under the corresponding fitted models. These empirical curves treat the sampled sources as complete and serve as illustrative summaries of the fitted source fluxes.

Before fitting the full \name\ model, which incorporates the optical-only sources into the population fit, we assess their compatibility  
with the population distribution inferred from the X-ray-only analysis via a posterior predictive check; details are provided in the Supplement. The results indicate that the observed X-ray counts of the optical-only sources fall within the predictive range implied by the X-ray-only model. This suggests that these sources occupy a region of the flux distribution where X-ray-detected sources are sparse but still present, supporting the use of the optical data to augment inference in the faint-intensity regime.

We first consider the double Pareto model. The full \name\ fit places the lower end of the population distribution in a fainter regime than the X-ray-only fit and yields a steeper faint-end \lns relation. The posterior distribution of the lower bound is more tightly constrained under the full \name\ fit, and the posterior distribution of the expected population size $\psi$ shifts toward larger values. This is consistent with the simulation study, where X-ray-only fits tended to place the lower bound at brighter values due to limited information about the faint end of the distribution.
The faint-end slope parameter $\theta_1$ also shifts toward larger values and is more tightly constrained under the full \name\ fit. Under (\ref{eq:lambda.prior}), this implies relatively greater probability mass near the lower end of the distribution and a steeper faint-end \lns relation.
The double Pareto model appears too restrictive for the CDF-S data. In the brighter, nearly X-ray-complete regime, the empirical \lns curves from the posterior source-flux samples are similar for the X-ray-only and \name\ fits. However, the corresponding population-level posterior median curves differ noticeably. This suggests that the double Pareto model does not provide enough flexibility to accommodate the additional faint-end information from the optically augmented sample while retaining the shape implied by the brighter sources. 
 
This motivates fitting a triple Pareto model, which introduces an additional low-intensity breakpoint and allows the faintest part of the population distribution to have its own power-law slope. We denote the slopes by $(\theta_0,\theta_1,\theta_2)$ and the knots by $(\tau_{S0},\tau_{S1},\tau_{S2})$, ordered from low to high flux (indexing the additional faint-end component by 0 to retain the double-Pareto labels for comparison). Thus, $\theta_0$ governs the faintest regime, while $\theta_1$ and $\theta_2$ describe the intermediate and bright-end regimes.
The triple Pareto fit resolves much of this tension. The \name\ and X-ray-only fits agree in the brighter, nearly X-ray-complete regime, where the X-ray detections already constrain the distribution well. The main difference occurs at the faint end.  
The full \name\ fit identifies an additional low-intensity breakpoint and a steep faint-end component, leading to a sharper increase in the inferred cumulative number of sources toward low fluxes. This structure is only weakly identified by the X-ray-only fit because of the limited number of X-ray detections in this part of the distribution, whereas the optically augmented sample provides the information needed to constrain it. 

\begin{figure}[t]
\begin{minipage}{\linewidth}
  \centering 
  \includegraphics[width=0.9\linewidth]{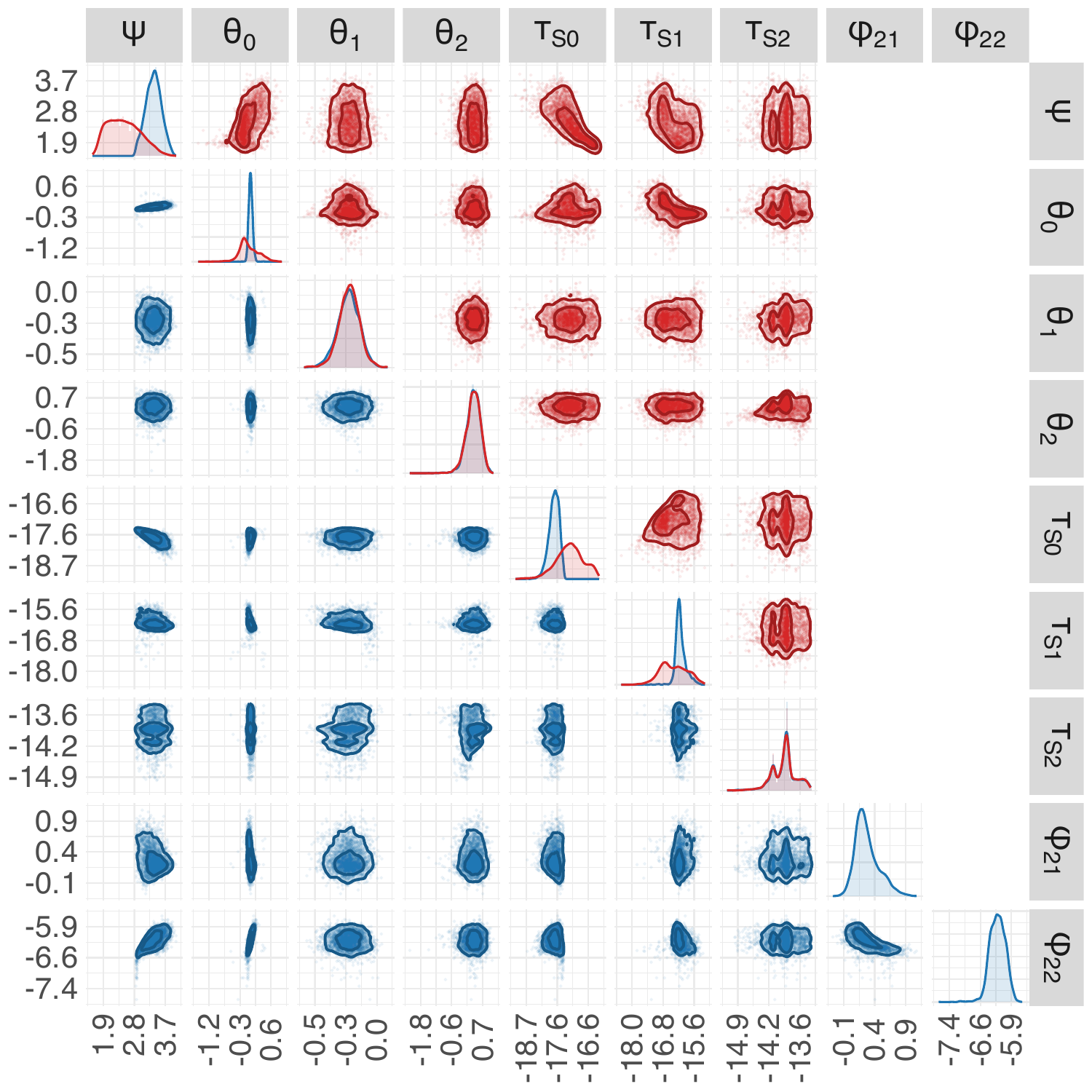} 
    \caption{
Pairwise marginal posterior distributions of the global model parameters for the CDF-S data under the triple Pareto model. The upper-right panels show the X-ray-only posterior in red, while the lower-left panels show the full \name\ posterior in blue. Marginal posterior densities are shown on the diagonal, and contours in the off-diagonal panels indicate $50\%$ and $90\%$ HPD regions. All parameter axes are displayed on a $\log_{10}$ scale. For visual consistency, the vertical axis in the top-left panel shows the parameter scale rather than the density scale. The optical incompleteness parameters $\phi_{21}$ and $\phi_{22}$ are included only in the full \name\ model. 
    \label{figure:pairwise_marginals_TP}}
  \end{minipage}
\end{figure}

Figure~\ref{figure:pairwise_marginals_TP} shows the pairwise marginal posterior distributions of the global model parameters under the triple Pareto model. The upper-right panels correspond to the X-ray-only fit, while the lower-left panels correspond to the full \name\ fit.  
Under the full \name\ fit, the posterior distributions of the lower knots $\tau_{S0}$ and $\tau_{S1}$ are substantially more tightly constrained than under the X-ray-only fit, and the faint-end slope parameter $\theta_0$ is much better  identified. This supports the interpretation from Figure~\ref{figure:LogN-LogS} that the optically augmented sample constrains an additional faint-end component of the \lns relation. By contrast, the posterior distributions of $\theta_1$, $\theta_2$, and the upper breakpoint $\tau_{S2}$ are similar under the X-ray-only and \name\ fits. These parameters are primarily informed by the brighter sources, for which X-ray detection is nearly complete, so the optically augmented sample adds comparatively little information in this regime.

The inferred population size $\psi$ is larger under the full \name\ model than under the X-ray-only model, reflecting the additional low-flux population implied by the optically augmented sample after correcting for incompleteness. 
The posterior distributions of the upper breakpoint $\tau_{S2}$ and the bright-end slope $\theta_2$ exhibit substantial uncertainty, a pattern seen in both the double and triple Pareto fits and under both the X-ray-only and full \name\ analyses. Notably, the marginal posterior distribution of $\tau_{S2}$ shows multiple modes, as illustrated in Figure~\ref{figure:pairwise_marginals_TP}. This reflects the few sources in the bright tail. In particular, the higher mode leaves the two brightest sources beyond the breakpoint, while the lower mode is associated with the next, moderately bright sources. Consequently, the corresponding slope parameter $\theta_2$ is only loosely constrained.
Consistent with this, the marginal posterior of $\theta_2$ overlaps substantially with high-probability regions of $\theta_1$, indicating limited evidence for a distinct upper breakpoint and separate bright-end power-law slope in these data. This is not unexpected for the CDF-S survey. As a deep-field survey with an extremely long exposure, the CDF-S region was selected to avoid very bright sources and to facilitate the detection of faint objects, and thus provides fewer constraints on the bright end of the piecewise-linear \lnsn. 
Finally, we consider the parameters governing the optical incompleteness function $g_{2\pmb w}(\lambda;\pmb\phi)$. The scale parameter $\phi_{22}$ is relatively well constrained, 
suggesting that the data identify the characteristic X-ray intensity scale over which the optical detection probability changes.  
In contrast, the shape parameter $\phi_{21}$ is less tightly constrained, suggesting that the data constrain the characteristic scale of the optical transition more strongly than its precise steepness. 

Overall, the CDF-S analysis shows the optically augmented \name\ fit provides information about the X-ray flux distribution beyond that available from the X-ray detections alone. 
After correcting jointly for X-ray and optical incompleteness, the full \name\ fit constrains an additional low-flux breakpoint and steep faint-end component that are only weakly identified in the X-ray-only analysis. 
At the same time, the triple Pareto model agrees with the X-ray-only fit in the brighter, nearly complete regime, indicating that the optical augmentation improves faint-end inference while preserving the fit in the brighter, nearly X-ray-complete regime. The flux scale of the inferred faint-end component is consistent with the regime in which deep CDF-S studies find a transition from AGN-dominated to galaxy-dominated X-ray source counts, 
based on source identification obtained from multiwavelength catalogue cross-matching \citep{lehmer20124,luo2017chandra}.  
Complementing these analyses, the full \name\ model incorporates the optical catalogue as an additional incomplete survey, so that optically selected source locations and their X-ray counts contribute to the joint likelihood and strengthen constraints on the faint-flux population where X-ray detections alone are sparse. Physically, this component may reflect the increasing contribution of normal or star-forming galaxies, including X-ray binaries, hot gas, and possibly weak or obscured nuclear accretion, although the present model does not distinguish between these source classes.


\section{Discussion}\label{sec:discussion}

In this paper, we develop \name, a novel method to combine incomplete (X-ray and optical) surveys for estimation of the galaxy \lns distribution in X-rays. Accurate modeling of \lns allows astronomers to answer questions about the geometry and structure of the universe. A major complication in the estimation of galaxy flux distributions in X-ray is large incompleteness in the low-intensity end of the distribution. Sources with low intensities are least likely to be detected, but most prevalent in the population. Ignoring the non-representativeness of the observed sources would lead to large inferential bias, and a potential misunderstanding of the underlying structure of the universe.

We demonstrate that optical survey data can be leveraged to substantially enlarge the observed sample at low source intensities, where X-ray detections are sparse. 
We carefully account for the incompleteness in X-ray-detected sources by incorporating an incompleteness function calibrated from previous survey simulations \citep{wright2023simulating}.
The incompleteness of the optically detected sources is modelled via a second (optical) incompleteness function with parameters fitted within the model. Evaluating our model on simulations realistic for the CDF-S data, we show that we can accurately recover \lns even under high proportions of incompleteness. 
The Poisson-binning formulation allows the incomplete source population to be modeled directly through counts in source-intensity and covariate bins, avoids explicit augmentation by the potentially large number of undetected sources, and facilitates computation, e.g., by allowing the incompleteness corrections to be largely precomputed.

In our CDF-S analysis, combining X-ray and optical survey data substantially strengthens inference at the faint end of the X-ray flux distribution. 
Under the triple-Pareto model, the \name\ fit constrains an additional low-flux breakpoint and steep faint-end component that are only weakly identified by the X-ray detections alone, with little change in the brighter, nearly X-ray-complete regime. This additional component may reflect an increasing contribution  of normal or star-forming galaxies at faint X-ray fluxes \citep{lehmer20124,luo2017chandra}.  

There are several natural extensions of the current analysis. 
There may be occurrences of sources that are detected in an optical catalog, having optical emission, but no X-ray emission (being X-ray \textit{dark}). Such sources would have zero probability of being detected in X-rays. To investigate such cases, a zero-inflated component could be added to the model, allowing source intensities to be zero.  
Allowing for such a component could refine inference at the faint end of the X-ray flux distribution. 
We will consider a zero-inflated model option in future work. 
Another major line of future work is the consideration of overlapping sources. Modelling each segment of overlap between sources would allow overlapping sources to be incorporated into the model, substantially increasing the available sample size and potentially improving posterior precision. This would, however, increase model complexity and the number of parameters, making model fitting more challenging. The current analysis uses the X-ray incompleteness function calibrated from survey simulations. Propagating uncertainty in this calibration through the population inference would be a useful extension. Other technical extensions include propagating uncertainty in the count-to-flux conversion, using more realistic source-specific PSFs, allowing more flexible spatial background models, and accounting for contributions of undetected sources to the background. 

Although our CDF-S analysis combines X-ray and optical survey data, the framework can readily incorporate an arbitrary number of surveys, e.g., additional optical, radio, or X-ray surveys, through their corresponding incompleteness functions. More generally, the Poisson-binning construction may be useful in other settings where a population is observed through multiple incomplete sampling mechanisms with different selection functions.

\begin{acks}[Acknowledgments]
In addition to the primary affiliation, MA also acknowledges affiliation with the Institute of Astronomy and Kavli Institute for Cosmology at the University of Cambridge, and an honorary affiliation with the Department of Mathematics at Imperial College London.
DvD and MA acknowledge partial support from the UK Engineering and Physical Sciences Research Council [EP/W015080/1, EP/W522673/1] and from the Marie-Skodowska-Curie RISE [H2020-MSCA-RISE-2019-873089] Grant provided by the European Commission. MA also acknowledges partial support from the European Union’s Horizon 2020 research and innovation programme under European Research Council Grant Agreement No 101002652 (BayeSN; PI K. Mandel). 
DCS acknowledges a Discovery Grant from The Natural Sciences and Engineering Research Council of Canada (RGPIN-2021-03985). DCS and MA acknowledge support from the Digital Research Alliance of Canada (\texttt{alliancecan.ca}).
AS and VLK acknowledge support from NASA Contract NAS8-03060 to the {\sl Chandra X-ray Center}.
This research has made use of data obtained from the Chandra Data Archive. 
This research also made use of software provided by the Chandra X-ray Center (CXC) in the application packages CIAO \citep{fruscione2006ciao, frucione2026ciao}, SAOImageDS9 \citep{fruscione2026ds9} and {\it Sherpa} \citep{freeman2001sherpa,siemiginowska2024sherpa}.
\end{acks}

\bibliographystyle{imsart-nameyear} 
\bibliography{bibliography}       

\newpage

\setcounter{section}{0}
\renewcommand\thesection{S\arabic{section}}

\setcounter{figure}{0}
\renewcommand\thefigure{S\arabic{figure}}

\setcounter{table}{0}
\renewcommand\thetable{S\arabic{table}}

\renewcommand{\theequation}{S.\arabic{equation}}
\renewcommand{\theHequation}{S\arabic{equation}}

{\huge\bf Supplement}


\section{Prior distributions}
In Table~\ref{tab:priors_final}, we list the prior distributions used for the triple-Pareto model in our CDF-S data analysis. The corresponding prior modifications for the double-Pareto model are described below.
\begin{table}[h!]
\centering
\caption{Prior distributions used for the triple Pareto model. Gamma distributions are parameterized in shape--rate form, such that $\mathbb{E}(X)=a/b$. }
\label{tab:priors_final}
\begin{tabular}{llll}
\hline
Parameter & Role & Prior & Hyperparameters \\
\hline

$\psi$ 
& Expected population size
& $\text{Gamma}(a_{\psi}, b_{\psi})$ 
& $a_{\psi}=1.5,\; b_{\psi}=10^{-3}$ \\

$\xi$ 
& Background count rate
& $\text{Gamma}(a_{\xi}, b_{\xi})$ 
& $a_{\xi}=1,\; b_{\xi}=10^{6}$ \\

$\theta_0$ 
& Faint-end slope
& $\text{Gamma}(a_{\theta_0}, b_{\theta_0})$ 
& $a_{\theta_0}=2,\; b_{\theta_0}=1$ \\

$\theta_1$ 
& Intermediate slope
& $\text{Gamma}(a_{\theta_1}, b_{\theta_1})$ 
& $a_{\theta_1}=2,\; b_{\theta_1}=1$ \\

$\theta_2$ 
& Bright-end slope
& $\text{Gamma}(a_{\theta_2}, b_{\theta_2})$ 
& $a_{\theta_2}=2,\; b_{\theta_2}=1$ \\

$\tau_0$ 
& Lower bound
& $\text{Gamma}(a_{\tau_0}, b_{\tau_0})$ (trunc.) 
& $a_{\tau_0}=1.5,\; b_{\tau_0}=10^{6}$ \\

$\tau_1$ 
& Faint breakpoint
& $\text{Gamma}(a_{\tau_1}, b_{\tau_1})$ (trunc.) 
& $a_{\tau_1}=1.1,\; b_{\tau_1}=10^{5}$ \\

$\tau_2 $ 
& Bright breakpoint
& $\text{TruncNorm}(\mu_{\tau_2}, \sigma_{\tau_2}^2)$ 
& see description \\

$\phi_{21}$ 
& Optical-detection shape
& $\text{Gamma}(a_{\phi_{21}}, b_{\phi_{21}})$ 
& $a_{\phi_{21}}=4,\; b_{\phi_{21}}=1$ \\

$\phi_{22}$ 
& Optical-detection scale
& $\text{Gamma}(a_{\phi_{22}}, b_{\phi_{22}})$ 
& $a_{\phi_{22}}=1.5,\; b_{\phi_{22}}=10^{6}$ \\

\hline
\end{tabular}
\end{table}

The prior on $\psi$ is chosen to be weakly informative, covering a wide
range of possible population sizes from nearly complete to strongly
incomplete regimes.
For the background rate $\xi$, we use a diffuse gamma prior that spans
realistic background levels encountered in X-ray observations.

In our analyses, we consider two- and three-component piecewise-linear \lns relationships. We first describe the prior specification for the three-component (triple Pareto) model, with slopes $(\theta_0,\theta_1,\theta_2)$ and ordered knots $(\tau_0<\tau_1<\tau_2)$, where $\tau_0$ is the lower bound of the population distribution and $\tau_1$ and $\tau_2$ are the two breakpoints. Following \cite{udaltsova2014universe} and \cite{stampoulis2017bayesian}, we specify gamma prior distributions on the power-law slopes that favour moderate
slopes while allowing sufficient flexibility. We specify truncated gamma and truncated Normal prior distributions on the lower bound and breakpoints, with truncation imposed to preserve their ordering.
The upper break parameter $\tau_2$ is constrained to lie in the bright-end
regime of the flux distribution, consistent with the expected location of the break at higher source intensities. Rather than specifying a conditional prior $\tau_2 \mid \tau_1$, we impose fixed truncation bounds derived from the empirical distribution of the naive $\hat{\lambda}_i$ estimates for the X-ray-detected sources, thereby decoupling $\tau_1$ and $\tau_2$ and improving sampler mixing. Specifically, the lower truncation bound and prior mean of the $\tau_2$ prior are both set to the empirical 75th percentile  
and the prior standard deviation to one half of the distance between the truncation bounds, where the upper truncation bound is set to the rightmost bin edge of
the  $\lambda$-grid, located in the far tail of the distribution where no extremely bright sources are observed. 
For the CDF-S data, the empirical 75th percentile is $5.27\times10^{-5}$ and the upper truncation bound is $0.004$, resulting in $\mu_{\tau_2}=5.27\times10^{-5}$ 
and $\sigma_{\tau_2}=1.97\times10^{-3}$. 
This results in a diffuse prior covering a broad range of plausible breakpoint locations.
Given this parametrisation, the prior on $\tau_1$ is specified as a gamma distribution with upper truncation at the lower truncation bound of $\tau_2$. This truncation occurs in the extreme right tail of the untruncated prior (e.g. above the 99th percentile under the chosen hyperparameters) and therefore has negligible influence on posterior inference. The fixed truncation provides a convenient way to preserve the ordering constraint while improving sampler stability.  
The fixed lower truncation bound for $\tau_0$ is set to the leftmost edge of the $\lambda$-grid, $5\times10^{-9}$, 
and its support is restricted to values below $\tau_1$, ensuring the required ordering $\tau_0<\tau_1$.

The parameters $\phi_{21}$ and $\phi_{22}$ governing the optical incompleteness function are assigned gamma priors. The prior on $\phi_{21}$ allows for a range of transition steepness comparable to those obtained in X-ray survey simulations \citep{wright2023simulating}. 
We assign $\phi_{22}$ a Gamma prior with the same hyperparameters as $\tau_0$ (and \(\tau_1\) for the double-Pareto model), providing a prior scale comparable to that of the lower-bound parameter of the X-ray flux distribution.

The double Pareto model is obtained by omitting the lowest-intensity component. Its parameters are \((\theta_1,\theta_2,\tau_1,\tau_2)\), with \(\tau_1\) representing the lower bound of the population distribution and \(\tau_2\) the single breakpoint. For the double Pareto model, we assign \(\tau_1\) the same prior used for \(\tau_0\) in the triple Pareto model, while the prior for \(\tau_2\) remains unchanged.
The slope parameters retain the same Gamma prior specification. For the simulation study, the numerical hyperparameter values and truncation constants were fixed to those obtained from the CDF-S analysis to ensure comparability across Monte Carlo runs.

\section{Metropolis-within-Gibbs sampler}\label{supp:mhg_sampler}
To obtain posterior inference, we implement a Metropolis-within-Gibbs sampler targeting~\eqref{form:full_posterior_with_cov}, with the population-level likelihood evaluated using the binned likelihood detailed in~\eqref{eq:loglik_reduced}.

The latent source intensities $\boldsymbol{\lambda}$ are updated component-wise using Metropolis--Hastings steps with proposals constructed on the log scale. For the triple-Pareto model, the proposals are truncated to respect the Pareto support constraint $\lambda_i>\tau_0$. The lower-bound parameter $\tau_0$ is updated using a Metropolis--Hastings step on the log scale, with an upper truncation at $\min\{\tau_1,\min_i\lambda_i\}$, ensuring that $\tau_0<\tau_1$ and $\tau_0<\lambda_i$ for all sources at each iteration. The breakpoint parameters $\tau_1$ and $\tau_2$ are updated individually using Metropolis--Hastings steps on the log scale, with truncation imposed as needed to preserve the ordering $\tau_0<\tau_1<\tau_2$ and the fixed prior truncation bounds described in the preceding section.

For the double-Pareto model, $\tau_1$ instead represents the lower bound of the population distribution. The latent-intensity proposals are therefore truncated to satisfy $\lambda_i>\tau_1$, and $\tau_1$ is updated using a Metropolis--Hastings step on the log scale with an upper truncation at $\min\{\tau_2,\min_i\lambda_i\}$. The breakpoint parameter $\tau_2$ is updated as in the triple-Pareto model.

The background rate $\xi$ is updated on the original scale using a normal Metropolis--Hastings proposal truncated below at zero. The population size parameter $\psi$ is updated via a Gibbs step using its gamma full conditional distribution. The slope parameters $\theta_j$ and the optical incompleteness parameters $(\phi_{21},\phi_{22})$ are updated individually using Metropolis--Hastings steps on the log scale. All Metropolis--Hastings acceptance probabilities account for the Jacobian induced by log-scale proposals and for any asymmetry introduced by truncated proposals.

Starting values for $\boldsymbol{\lambda}$ and $\xi$ are obtained from naive estimates based on the observed counts, with the $\boldsymbol{\lambda}$ starting values adjusted if needed to satisfy positivity and the corresponding Pareto support constraint. All remaining parameters are initialized by sampling from their respective prior distributions, subject to the corresponding support constraints. Proposal standard deviations are tuned during the run using a fixed schedule (10,000, 20,000, 40,000, 60,000, and 100,000 iterations), based on the empirical standard deviation of the draws within each adaptation window. To ensure that all adaptation occurs prior to inference, the first 200,000 iterations are discarded as burn-in.


\section{X-ray Detection Incompleteness and Its Numerical Evaluation}\label{sec:incompleteness}

\subsection{Bilinear interpolation of the Wright incompleteness model}

We perform bilinear interpolation at the level of the X-ray detection probability, rather than interpolating the tabulated parameters \(\nu_1\) and \(\nu_2\). The interpolation is specified explicitly below. 
Let the calibration grid consist of background rates
$\xi_1 < \cdots < \xi_I$
and off-axis angles
$L_1 < \cdots < L_J$.
Suppose $(\xi^\ast,L^\ast)$ lies in the grid cell satisfying
\[
\xi_i \le \xi^\ast \le \xi_{i+1},
\qquad
L_j \le L^\ast \le L_{j+1}.
\]
Define interpolation weights
\[
t = \frac{\xi^\ast - \xi_i}{\xi_{i+1} - \xi_i},
\qquad
u = \frac{L^\ast - L_j}{L_{j+1} - L_j},
\qquad
t,u \in [0,1].
\]

At the four surrounding grid nodes
$(\xi_{i+a-1},L_{j+b-1})$ with $a,b\in\{1,2\}$, the model of \cite{wright2023simulating} provides exponents
\[
\nu_1^{(ab)}=\nu_1(\xi_{i+a-1}  \mathcal{T},L_{j+b-1}),
\qquad
\nu_2^{(ab)}=\nu_2(\xi_{i+a-1}  \mathcal{T},L_{j+b-1}),
\]
yielding detection probabilities
\[
g_{ab}(\lambda_i)
=
1-\exp\!\left(
-\frac{(\lambda_i \mathcal{T} e_i)^{\nu_1^{(ab)}}}
{10^{\nu_2^{(ab)}}}
\right).
\]

\noindent
The interpolated detection probability is then 
\begin{align}
\label{eq:bilinear_interp}
g_{1}(\lambda_i \mid \xi^\ast  \mathcal{T},L^\ast,e_i)
&=
(1-t)(1-u)\, g_{11}(\lambda_i)
+ (1-t)u\, g_{12}(\lambda_i) \\
&\quad
+ t(1-u)\, g_{21}(\lambda_i)
+ t u\, g_{22}(\lambda_i). \nonumber
\end{align}

\subsection{Efficient incompleteness correction via Poisson binning}
\label{supp:precomputation_details}

We describe in detail how the Poisson-binning formulation of Section~\ref{section:binning_multiple_surveys} enables efficient precomputation of (i) source-specific values of the X-ray incompleteness function $g_{1\pmb w}(\lambda)$ via a lookup table, and (ii) Monte Carlo (MC) approximations of the $\pmb W$-marginal terms $\hat\pi_j$ appearing in the binned likelihood \eqref{ref:likelihood_binned_multiple}. Throughout, we consider the two-survey setting with sequential detection probabilities defined in \eqref{eq:complete}, where $j=1$ denotes X-ray detections and $j=2$ optical-only detections.

\subsubsection{Precomputation of a lookup table for source-specific incompleteness corrections}

Let \(\{\mathcal B_k\}_{k=1}^B\) denote the intensity bins introduced in Section~\ref{section:binning_one_survey}, with midpoints \(b_k\). In the Poisson-binned likelihood \eqref{ref:likelihood_binned_multiple}, the dependence of the incompleteness functions on \(\lambda\) is therefore restricted to the fixed grid \(\{b_k\}\), while their dependence on \(\pmb W\) is represented through the \(\pmb W\)-bin representatives \(\pmb c_\ell\). 

\begin{figure}[t]
\centering
\begin{subfigure}{0.53\linewidth}
    \centering    
    \includegraphics[width=\linewidth]{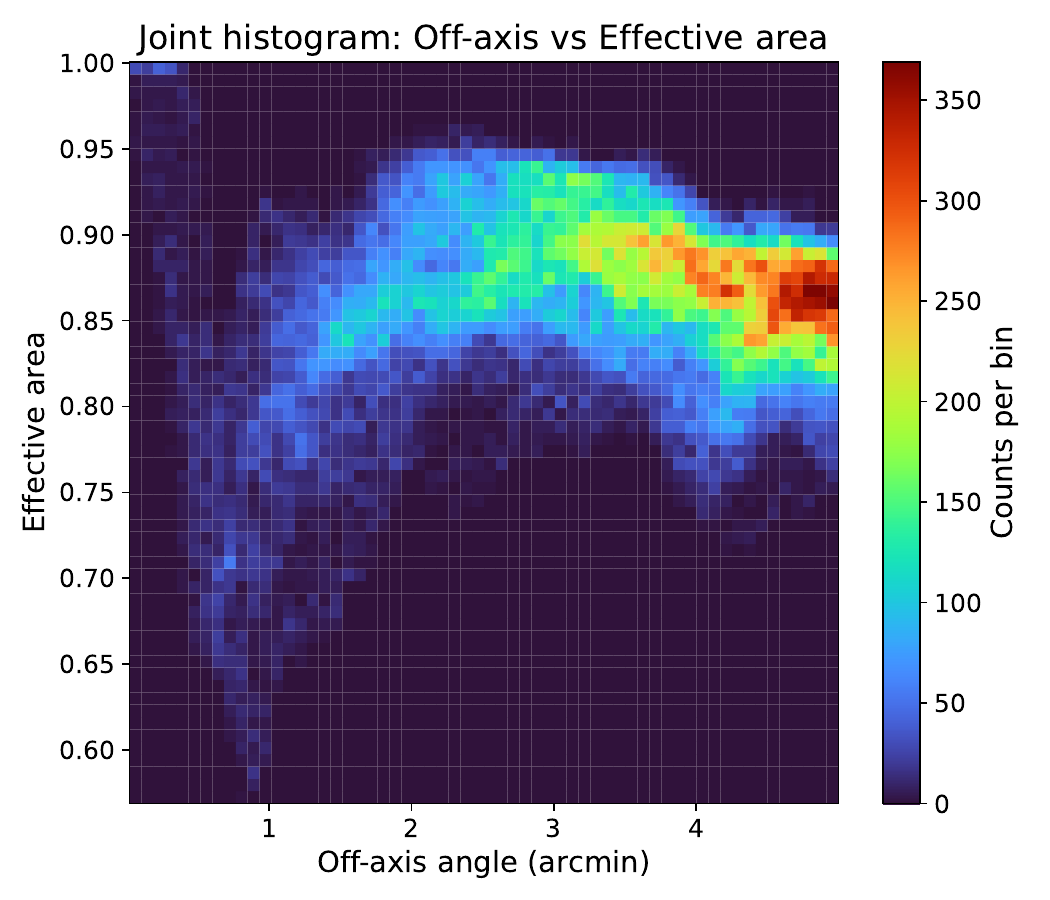} 
    \caption{Normalized effective area $e$ vs. off-axis  angle $L$.}
    \label{figure:joint_hist_E_L}
\end{subfigure}
\begin{subfigure}{0.35\linewidth}
    \centering
    \includegraphics[width=\linewidth]{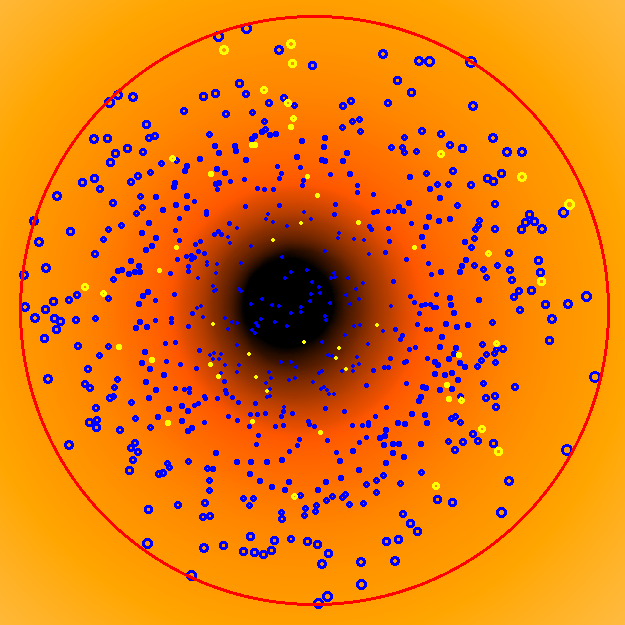}
    \medskip
    \caption{CDF-S PSF map.}
    \label{figure:psf_map}
\end{subfigure}
\hfill
\caption{(a) Joint histogram of normalized effective area $e$ and off-axis angle $L$ based on
$ M = 100,000$  locations drawn uniformly over the analyzed CDF-S region. The figure shows the dependence of effective area on position in the field, with typically lower values at larger off-axis angles and additional spatial variation due to the structure of the exposure map. 
(b)  PSF map of the CDF-S subregion $\mathcal{S}$. The PSF width increases with off-axis angle, reflecting the degradation of angular resolution away from the aimpoint. Circles indicate source locations, with X-ray-detected sources shown in yellow and optical-only sources in blue. The PSF map is used to define circular source regions whose sizes depend on position in the field.}
\label{figure:supp_psf_and_joint_EL} 
\end{figure}

In our implementation, we choose the $\pmb W$-binning to be extremely fine by
approximating $p_{\pmb W}$ using a Monte Carlo sample of $M=100{,}000$ locations
drawn uniformly over the analyzed CDF-S region, as described in
Section~\ref{sec:implement},
\[
\pmb W^{(\ell)}=(L^{(\ell)},e^{(\ell)}), \qquad \ell=1,\ldots,M.
\]
For the \(\pmb W\)-marginalization,  the Monte Carlo approximation can be interpreted as inducing a sufficiently
fine $\pmb W$-binning that approximates the distribution $p_{\pmb W}$, with equal bin
weights $\omega_\ell=1/M$. 
At this resolution, each observed source can be treated as occupying its own $W$-cell, with the representative point for the observed-source contribution to the likelihood chosen to be the observed covariate value $W_i=(L_i,e_i)$.  
This justifies evaluating the
source-specific incompleteness terms in~\eqref{ref:likelihood_binned_multiple} at the observed $\pmb W_i$, while
the Monte Carlo sample is used to approximate the $\pmb W$-marginalization in the 
normalizing terms $\hat\pi_j$ through the corresponding finite sums.
The benefit of this fine discretization is particularly transparent for the normalized effective area \(e\). Since \(e\) enters the X-ray detection function multiplicatively with the source intensity \(\lambda\), approximation error in \(e\) translates directly into a corresponding proportional shift in the effective intensity scale entering the detection probability. 
Figure~\ref{figure:supp_psf_and_joint_EL} shows the joint histogram of normalized effective area $e$ and off-axis
angle $L$, illustrating how the effective area varies with off-axis position
across the field.

Under the Poisson-binning model
\eqref{eq:bincounts}--\eqref{form:X_k_Poisson_multiple}, the observed bin
counts are independent Poisson random variables, and the corresponding
log-likelihood for two surveys, up to an additive constant, is
\begin{align}
l^*(\psi,\pmb\theta,\pmb\phi)
=
\sum_{j=1}^2\sum_{k=1}^B\sum_{\ell=1}^M
X_{jk\ell}
\log\!\Big(
\psi\,g_{j\pmb c_\ell}(b_k)\,
p_\Lambda(b_k\mid\pmb\theta)\,
\omega_\ell
\Big)
-
\psi\sum_{j=1}^2
\hat\pi_j(\pmb\theta,\pmb\phi),
\label{eq:loglik_binned_full}
\end{align}
where
\begin{align}
\hat\pi_j(\pmb\theta,\pmb\phi)
=
\sum_{k=1}^B\sum_{\ell=1}^M
g_{j\pmb c_\ell}(b_k)\,
p_\Lambda(b_k\mid\pmb\theta)\,
\delta_k\,\omega_\ell,
\qquad j\in\{1,2\}.
\label{eq:pihat_def_discrete}
\end{align}
Thus, computation requires (a) evaluating
$g_{j\pmb c_\ell}(b_k)$ at many $(k,\ell)$ pairs and
(b) repeatedly evaluating the sums in
\eqref{eq:pihat_def_discrete} as
$(\pmb\theta,\pmb\phi)$ change during MCMC.
Direct evaluation of these quantities within each MCMC iteration would be
computationally prohibitive. The binning structure allows the expensive
components of these calculations to be precomputed, leaving only relatively
inexpensive bin-wise evaluations within the sampler.

Given the fine $\pmb W$-binning described above and the fixed $\lambda$-grid
$\{b_k\}$, we precompute the X-ray detection probability for each observed
source and each intensity bin:
\begin{align}
G_{ik}
:=
g_{1\pmb W_i}(b_k)
=
g_1(b_k;\pmb W_i,\hat{\xi}),
\qquad
i=1,\ldots,n,\quad k=1,\ldots,B,
\label{eq:g_lookup}
\end{align}
where $\hat{\xi}=X/(A\mathcal{T})$ is the empirical background-rate estimate
used for evaluation of the X-ray incompleteness function, as described in
Section~\ref{section:incompleteness_marginalization}. The matrix $G=(G_{ik})$ is computed once for a fixed bin grid
$\{b_k\}$ and fixed $\hat{\xi}$ and is then used as a lookup table inside the
sampler. This removes the need for repeated bilinear interpolation and
evaluation of the X-ray incompleteness function within the sampler,
replacing both with a simple lookup-table evaluation.

\subsubsection{Monte Carlo factors for $\pmb W$-marginalization}

The normalizing terms $\hat\pi_j$ require marginalization over $p_{\pmb W}$. 
Using the Monte Carlo sample $\{\pmb W^{(\ell)}\}_{\ell=1}^M$, we compute for each intensity bin $k$
\begin{align}
\bar g_{1,k}
&:=
\frac{1}{M}\sum_{\ell=1}^M
g_{1\pmb W^{(\ell)}}(b_k;\hat\xi),
\label{supp:gbar}
\\
\bar q_k
&:=
\frac{1}{M}\sum_{\ell=1}^M
\left(1-g_{1\pmb W^{(\ell)}}(b_k;\hat\xi)\right)
=
1-\bar g_{1,k}.
\label{supp:qbar}
\end{align}
These quantities represent Monte Carlo bin-wise averages of the X-ray detection probability and its complement. Because Poisson binning restricts the $\lambda$-dependence to the fixed grid $\{b_k\}$, these Monte Carlo averages can be precomputed once for each intensity bin rather than recomputed at every sampling iteration.

Because the second factor of the optical incompleteness function, which we denote by $h(b_k;\pmb\phi_2)$, depends only on $\lambda$ and $\pmb\phi_2$, it does not vary across the $\pmb W$-bins. Specifically,
\[
h(b_k;\pmb\phi_2)
=
1-\exp\!\left[
-\left(\frac{b_k}{\phi_{22}}\right)^{\phi_{21}}
\right].
\]
Under the Monte Carlo approximation, the corresponding bin-wise average of the optical-only detection probability is therefore
\[
\frac{1}{M}\sum_{\ell=1}^M
g_{2\pmb W^{(\ell)}}(b_k;\pmb\phi_2)
=
\bar q_k\,h(b_k;\pmb\phi_2).
\]

For the population component of the normalizing terms, we compute the probability mass of each intensity bin exactly under the Pareto-mixture model \eqref{eq:lambda.prior}. Let $F_\Lambda(\cdot\mid\pmb\theta)$ denote its CDF and let $u_0,\ldots,u_B$ denote the corresponding bin edges. For finite bins,
\[
m_k(\pmb\theta)
=
F_\Lambda(u_k\mid\pmb\theta)
-
F_\Lambda(u_{k-1}\mid\pmb\theta),
\]
and, if a tail bin is included,
\[
m_{B+1}(\pmb\theta)
=
1-F_\Lambda(u_B\mid\pmb\theta).
\]

Combining the Monte Carlo approximations \eqref{supp:gbar}--\eqref{supp:qbar} with the exact Pareto-mixture bin masses yields
\begin{align}
\hat\pi_1(\pmb\theta)
&\approx
\sum_{k=1}^{B+1}
\bar g_{1,k}\,
m_k(\pmb\theta),
\label{eq:pi1_hat_final}
\\
\hat\pi_2(\pmb\theta,\pmb\phi_2)
&\approx
\sum_{k=1}^{B+1}
\bar q_k\,
h(b_k;\pmb\phi_2)\,
m_k(\pmb\theta).
\label{eq:pi2_hat_final}
\end{align}

\subsubsection{Computational form of the binned likelihood}

The preceding subsections introduced two sets of precomputed quantities, the source-specific lookup matrix $G_{ik}$ and the Monte Carlo factors $(\bar g_{1,k},\bar q_k)$ used for $\pmb W$-marginalization. We now show how these quantities reduce the general binned likelihood \eqref{eq:loglik_binned_full} to the computational form evaluated within the sampler. In particular, repeated evaluation of the X-ray incompleteness function and explicit summation over $\pmb W$-cells are replaced by lookup-table evaluations and precomputed Monte Carlo averages.

Let $k(i)$ denote the intensity bin containing the current value of $\lambda_i$. Using the exact bin masses $m_k(\pmb\theta)$ defined above for the normalizing terms, the log-likelihood evaluated inside the sampler becomes
\begin{align}
\label{eq:loglik_reduced}
l^*(\psi,\pmb\theta,\pmb\phi_2)
&=
\sum_{i=1}^{n}
\log\!\Big(
\psi\,\tilde g_i\,p_\Lambda(b_{k(i)}\mid\pmb\theta)
\Big)
\nonumber\\
&\quad
-\psi\left[
\sum_{k=1}^{B+1} \bar g_{1,k}\,m_k(\pmb\theta)
+
\sum_{k=1}^{B+1}
\bar q_k\,h(b_k;\pmb\phi_2)\,m_k(\pmb\theta)
\right]
+c,
\end{align}
where
\[
\tilde g_i
=
\begin{cases}
G_{i,k(i)}, & \text{if } \mathcal J_i=1,\\[4pt]
(1-G_{i,k(i)})\,h(b_{k(i)};\pmb\phi_2), & \text{if } \mathcal J_i=2,
\end{cases}
\]
and $c$ collects terms constant in $(\psi,\pmb\theta,\pmb\phi_2)$.

In summary, the computational savings arise from Poisson binning in $\lambda$, together with the fine Monte Carlo representation of the covariate distribution $p_{\pmb W}$. Binning in $\lambda$ restricts the incompleteness-function evaluations to the fixed grid $\{b_k\}$, allowing the source-specific X-ray detection probabilities to be precomputed in the lookup matrix $G_{ik}$. The same fixed grid, combined with the Monte Carlo representation of $p_{\pmb W}$, allows the $\pmb W$-marginalization in the normalizing terms to be precomputed through the bin-wise averages $(\bar g_{1,k},\bar q_k)$. Consequently, evaluation of \eqref{eq:loglik_reduced} within the sampler requires neither repeated bilinear interpolation and evaluation of the X-ray incompleteness function nor repeated Monte Carlo averaging over $\pmb W$, thereby making posterior sampling computationally feasible.


\clearpage

\section{Additional Results for the Simulation Study and CDF-S Data Analysis}\label{section:supp_additional_cdfs_results}

\subsection{Additional simulation-study results}
In addition to the coverage and posterior interval-width summaries shown in Figure~\ref{fig:coverage_simulations}, we assess point-estimation accuracy using the root mean squared error (RMSE) of the posterior mean. Table~\ref{tab:simulation_rmse} reports the resulting RMSEs for the \name\ and X-ray-only analyses.  Since the parameter values vary across simulation replicates, the reported RMSEs are computed using the posterior mean and true value from each replicate.

\begin{table}[h!]
\centering
\caption{Root mean squared error (RMSE) of the posterior mean across the 100 simulation replicates. The final column gives the ratio of the X-ray-only RMSE to the \name\ RMSE. Values greater than one indicate lower estimation error under \name. The optical incompleteness parameters $\phi_{21}$ and $\phi_{22}$ are not present in the X-ray-only model.}
\label{tab:simulation_rmse}
\begin{tabular}{lccc}
\hline
Parameter & HELIOS & X-ray only & RMSE ratio \\
\hline
$\xi$       & $2.25\times10^{-10}$ & $2.29\times10^{-10}$ & 1.02 \\
$\psi$      & $8.09\times10^{2}$   & $1.22\times10^{3}$   & \textbf{1.51} \\
$\theta_1$  & $6.71\times10^{-2}$  & $1.05\times10^{-1}$  & \textbf{1.56} \\
$\theta_2$  & $1.40$                & $1.40$                & 1.00 \\
$\tau_1$    & $7.04\times10^{-8}$  & $5.95\times10^{-7}$  & \textbf{8.44} \\
$\tau_2$    & $7.70\times10^{-4}$  & $7.86\times10^{-4}$  & 1.02 \\
$\phi_{21}$ & $1.76$                & --                    & -- \\
$\phi_{22}$ & $1.96\times10^{-7}$  & --                    & -- \\
\hline
\end{tabular}
\end{table}

The RMSE results are consistent with the posterior interval-width comparisons in Figure~\ref{fig:coverage_simulations}. Incorporating the optical-only sources leads to substantially improved estimation of the faint-end parameters, with the largest reduction in RMSE occurring for the lower bound $\tau_1$. The RMSE is also reduced for $\psi$ and $\theta_1$, whereas the results for $\xi$, $\theta_2$, and $\tau_2$ are very similar under the two analyses.

\subsection{Compatibility of the optical-only sources with the X-ray-only CDF-S fit}
\label{section:supp_posterior_predictive}

To assess whether the optical-only sources are compatible with the population model inferred from the X-ray data alone, we carry out an out-of-sample posterior predictive check based on the X-ray-only fit. The goal of this analysis is to determine whether the observed X-ray counts of the optical-only sources fall in a plausible range under the X-ray-only fitted model, rather than representing a population of sources that is systematically much fainter than would be supported by the X-ray-detected sample.

For each retained posterior draw of $(\xi,\pmb\theta)$ from the X-ray-only analysis, and for each optical-only source with observed covariates $\pmb W_i$, we first generate a latent intensity from the population distribution conditional on the source not having been detected in X-rays,
\[
p(\lambda_i \mid Z_{i1}=0,\pmb W_i,\hat{\xi},\pmb\theta)
\propto
\bigl(1-g_{1\pmb W_i}(\lambda_i;\hat{\xi})\bigr)\,
p_\Lambda(\lambda_i\mid\pmb\theta).
\]
We sample from this conditional distribution by rejection sampling, using proposals from $p_\Lambda(\cdot\mid\pmb\theta)$ and acceptance probability $1-g_{1\pmb W_i}(\lambda_i;\hat{\xi})$. Conditional on the accepted latent intensities, replicated X-ray counts are then generated from the object-level model in \eqref{eq:observed_model}, using the corresponding posterior draw of $\xi$. Repeating this procedure across posterior draws yields a posterior predictive distribution for the X-ray counts of each optical-only source under the X-ray-only fitted model.

\begin{figure}[t]
    \centering
    \includegraphics[width=0.75\linewidth]{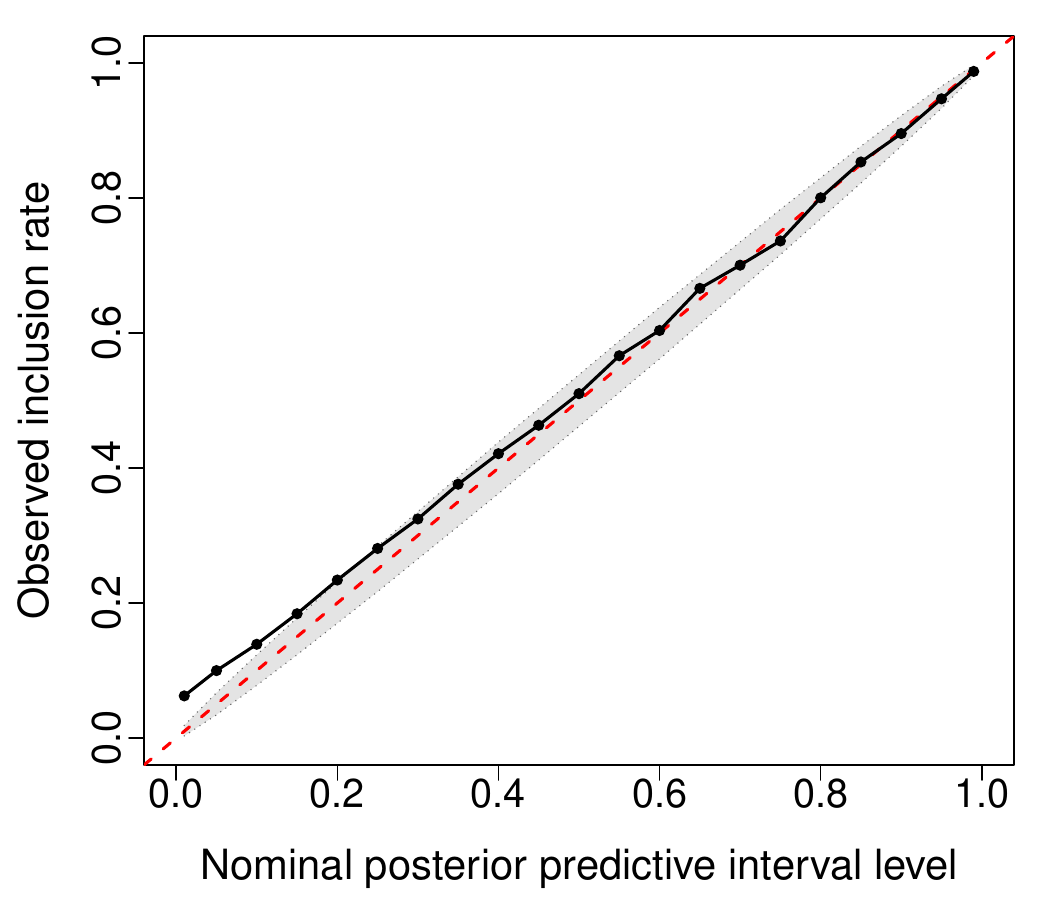}
    \caption{
Posterior predictive interval inclusion plot for the X-ray counts of the optical-only sources under the X-ray-only fitted model. For each nominal posterior predictive interval probability $\alpha$, the vertical axis shows the empirical proportion of observed X-ray counts contained within the corresponding equal-tailed posterior predictive intervals. The dashed diagonal line gives the nominal reference, and the shaded band indicates the range of inclusion probabilities expected from finite-sample variability.   
\label{figure:posterior_predictive_xr_to_optical}}
\end{figure}

To evaluate the agreement between the observed X-ray counts of the optical-only sources and this posterior predictive distribution, we compute equal-tailed posterior predictive intervals for each source over a grid of nominal levels $\alpha$, and record the empirical proportion of observed counts falling within the corresponding intervals. Figure~\ref{figure:posterior_predictive_xr_to_optical} shows that the empirical inclusion proportions track the nominal posterior predictive interval levels closely, with slightly higher inclusion rates at some values of $\alpha$. Thus, the optical-only sources are broadly contained within the predictive range implied by the X-ray-only fit. In particular, the results provide no indication that the optical-only sample is concentrated in a regime of source intensities far below that supported by the X-ray-detected sample alone. Rather, the optical-only sources appear to extend the information available from the X-ray sample into a region where X-ray-detected sources are sparse, but still present, supporting the use of the optical sources to augment inference in this lower-intensity regime.

\subsection{Additional posterior summaries and diagnostics for the CDF-S analysis}

\subsubsection{Posterior summaries}
Tables~\ref{tab:triple_pareto_posterior_comparison} and
\ref{tab:double_pareto_posterior_comparison} report posterior medians and
95\% credible intervals for the triple- and double-Pareto CDF-S analyses,
respectively. Across both population models, the largest differences between
the \name\ and X-ray-only fits occur in parameters governing the faint end of
the source-intensity distribution and in the population size $\psi$, while the
bright-end parameters $\theta_2$ and $\tau_2$ are very similar under the two
fits. In the triple-Pareto analysis, the inclusion of the optical-only sources
substantially reduces uncertainty in the faint-end slope $\theta_0$ and lower bound $\tau_0$. 
In the double-Pareto analysis, the corresponding differences are
most apparent for $\theta_1$ and $\tau_1$.

\begin{table}[t]
\centering
\caption{Posterior summaries for the triple-Pareto CDF-S analysis. Reported are posterior medians and 95\% credible intervals for the \name\ and X-ray-only fits.}
\label{tab:triple_pareto_posterior_comparison}
\begin{tabular}{lcccc}
\toprule
& \multicolumn{2}{c}{\name} & \multicolumn{2}{c}{X-ray only} \\
\cmidrule(lr){2-3} \cmidrule(lr){4-5}
Parameter & Median & 95\% CI & Median & 95\% CI \\
\midrule
$\xi$
& $1.64\times10^{-7}$
& $(1.64\times10^{-7},\,1.65\times10^{-7})$
& $1.64\times10^{-7}$
& $(1.64\times10^{-7},\,1.65\times10^{-7})$ \\

$\psi$
& $2105$
& $(783,\,5598)$
& $273$
& $(60,\,2547)$ \\

$\theta_0$
& $1.03$
& $(0.82,\,1.35)$
& $0.81$
& $(0.28,\,3.70)$ \\

$\theta_1$
& $0.58$
& $(0.40,\,0.81)$
& $0.59$
& $(0.41,\,0.80)$ \\

$\theta_2$
& $2.10$
& $(0.48,\,6.13)$
& $2.10$
& $(0.48,\,6.14)$ \\

$\tau_0$
& $1.46\times10^{-7}$
& $(5.65\times10^{-8},\,2.81\times10^{-7})$
& $5.25\times10^{-7}$
& $(4.81\times10^{-8},\,3.92\times10^{-6})$ \\

$\tau_1$
& $5.80\times10^{-6}$
& $(3.18\times10^{-6},\,1.98\times10^{-5})$
& $3.30\times10^{-6}$
& $(4.31\times10^{-7},\,3.14\times10^{-5})$ \\

$\tau_2$
& $1.09\times10^{-3}$
& $(2.89\times10^{-4},\,3.33\times10^{-3})$
& $1.10\times10^{-3}$
& $(2.94\times10^{-4},\,3.34\times10^{-3})$ \\

$\phi_{21}$
& $1.92$
& $(1.04,\,6.20)$
& -- & -- \\

$\phi_{22}$
& $6.31\times10^{-7}$
& $(2.75\times10^{-7},\,1.41\times10^{-6})$
& -- & -- \\
\hline 
\end{tabular}
\end{table}

\begin{table}[t]
\centering
\caption{Posterior summaries for the double-Pareto CDF-S analysis. Reported are posterior medians and 95\% credible intervals for the \name\ and X-ray-only fits.}
\label{tab:double_pareto_posterior_comparison}
\begin{tabular}{lcccc}
\toprule
& \multicolumn{2}{c}{\name} & \multicolumn{2}{c}{X-ray only} \\
\cmidrule(lr){2-3} \cmidrule(lr){4-5}
Parameter & Median & 95\% CI & Median & 95\% CI \\
\midrule
$\xi$
& $1.64\times10^{-7}$
& $(1.64\times10^{-7},\,1.65\times10^{-7})$
& $1.640\times10^{-7}$
& $(1.64\times10^{-7},\,1.65\times10^{-7})$ \\

$\psi$
& $1223$
& $(677,\,3516)$
& $170$
& $(58,\,1064)$ \\

$\theta_1$
& $0.84$
& $(0.73,\,0.98)$
& $0.60$
& $(0.45,\,0.78)$ \\

$\theta_2$
& $2.03$
& $(0.40,\,6.08)$
& $2.11$
& $(0.47,\,6.12)$ \\

$\tau_1$
& $1.50\times10^{-7}$
& $(4.50\times10^{-8},\,2.77\times10^{-7})$
& $6.16\times10^{-7}$
& $(3.42\times10^{-8},\,3.84\times10^{-6})$ \\

$\tau_2$
& $1.18\times10^{-3}$
& $(2.15\times10^{-4},\,3.54\times10^{-3})$
& $1.11\times10^{-3}$
& $(2.93\times10^{-4},\,3.37\times10^{-3})$ \\

$\phi_{21}$
& $2.87$
& $(1.06,\,7.84)$
& -- & -- \\

$\phi_{22}$
& $3.49\times10^{-7}$
& $(8.47\times10^{-8},\,7.10\times10^{-7})$
& -- & -- \\
\bottomrule
\end{tabular}
\end{table}

\begin{figure}[t]
    \centering
    \includegraphics[width=0.75\linewidth]{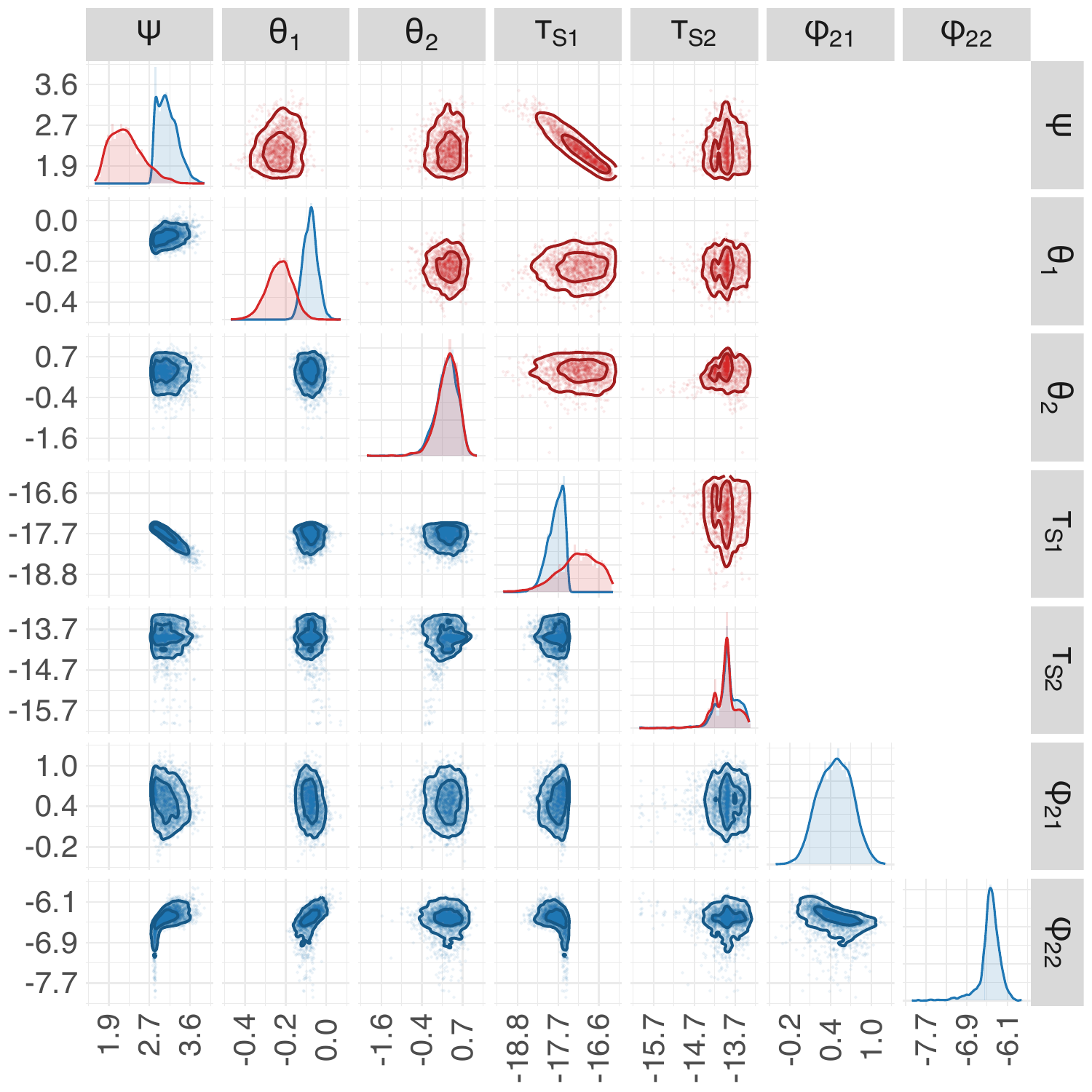}
    \caption{Pairwise posterior distributions for the double-Pareto CDF-S analysis under the \name\ and X-ray-only fits. Marginal posterior densities are shown along the diagonal and pairwise posterior contours off the diagonal. Parameters are shown on the $\log_{10}$ scale. Blue represents the \name\ fit and red the X-ray-only fit. \label{figure:pairwise_marginals_DP}}
\end{figure}

Figure~\ref{figure:pairwise_marginals_DP} shows the corresponding marginal
and pairwise posterior distributions for the double-Pareto model. The figure
illustrates both the changes in the faint-end posterior distributions between
the \name\ and X-ray-only analyses and the posterior dependence among the
population parameters.

\begin{figure}[h!]
\centering 
\begin{minipage}[t]{.99\linewidth}
  \centering
    \includegraphics[width=0.99\linewidth]{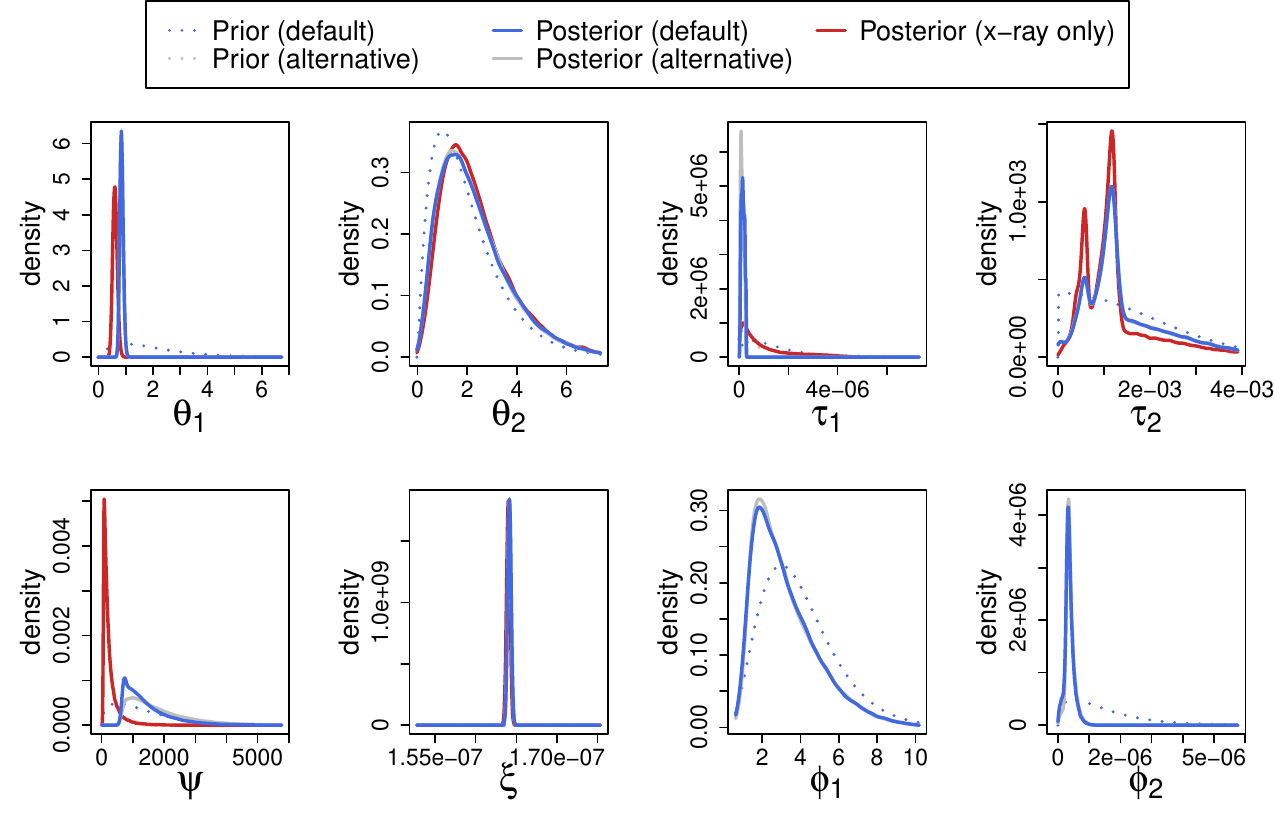}
    \caption{Prior sensitivity analysis for the double-Pareto CDF-S fit. Marginal posterior distributions under the default and alternative priors for $\tau_1$ are shown in blue and gray, respectively, with the corresponding prior densities shown by dotted lines. The X-ray-only posterior is shown in red for comparison. The alternative prior $\mathrm{Gamma}(0.6,3\times10^5)$ places substantially more mass on smaller values of $\tau_1$, closer to zero, than the default $\mathrm{Gamma}(1.5,10^6)$ prior. All parameters are shown on their original scale.  
    \label{figure:marginals_DP_origscale}  } 
\end{minipage}  
\end{figure}

\paragraph{Prior sensitivity}
To assess sensitivity to the prior on the lower bound $\tau_1$ in the double-Pareto model, we repeated the analysis using a $\mathrm{Gamma}(0.6,3\times10^5)$ prior in place of the default $\mathrm{Gamma}(1.5,10^6)$ prior. The alternative prior is more diffuse and places substantially more mass on smaller values of $\tau_1$, closer to zero. This allows us to test whether the posterior for $\tau_1$ is being held away from zero by the default prior. Figure~\ref{figure:marginals_DP_origscale} shows that the posterior distribution of $\tau_1$ changes little under the alternative prior, while the remaining parameters are unchanged or only mildly affected. This indicates that the faint-source data are sufficiently informative to constrain the lower bound and that the result is not primarily driven by the default prior specification.

\clearpage
\subsubsection{Posterior sampling diagnostics}
\begin{figure}[t]
    \centering
    \includegraphics[width=0.99\linewidth]{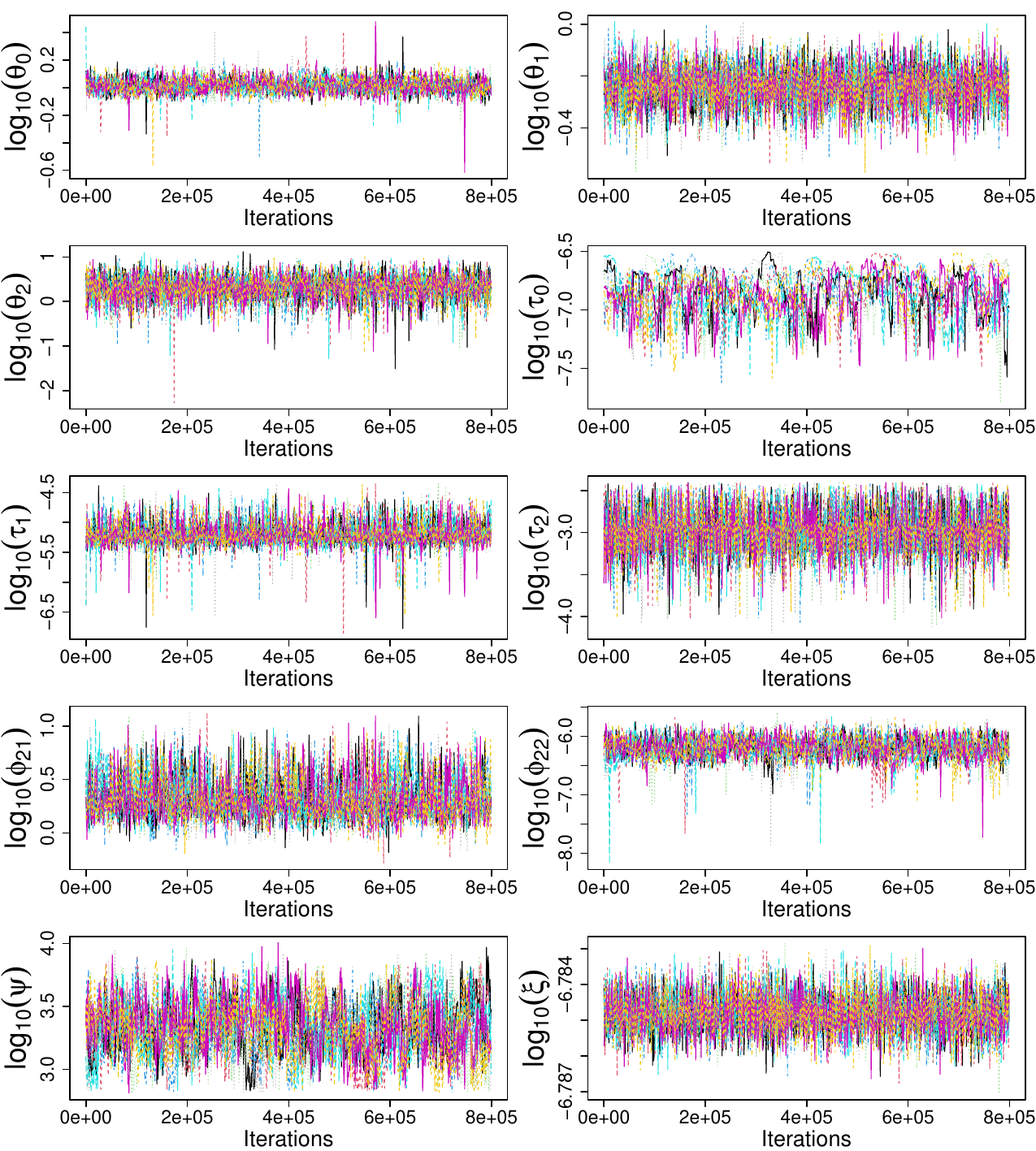}
    \caption{Traceplots for the eight MCMC chains of the Metropolis-within-Gibbs sampler for the triple-Pareto \name\ CDF-S analysis. Parameters are shown on the $\log_{10}$ scale, with every 1000th post-burn-in iteration displayed. Visual inspection indicates satisfactory mixing across chains, with no obvious convergence issues. 
    \label{figure:trace_CDFS_TP}} 
\end{figure}

\FloatBarrier
Tables~\ref{tab:triple_pareto_diagnostics} and
\ref{tab:double_pareto_diagnostics} report Gelman--Rubin $\hat R$ statistics
and effective sample sizes (ESS) for the triple- and double-Pareto analyses,
respectively. For the triple-Pareto fits, all $\hat R$ values are at most
$1.009$. The smallest ESS is $512$ for $\tau_0$ in the \name\ fit, reflecting
slower mixing for this parameter, while the remaining effective sample sizes
are generally substantially larger. Figure~\ref{figure:trace_CDFS_TP} shows
representative traceplots for the eight chains of the triple-Pareto \name\
analysis and provides no indication of chain separation or systematic drift.

\begin{table}[h]
\centering
\caption{MCMC diagnostics for the triple-Pareto CDF-S analysis. Reported are $\hat R$ and effective sample size (ESS) for the \name\ and X-ray-only fits.}
\label{tab:triple_pareto_diagnostics}
\begin{tabular}{lrrrr}
\toprule
& \multicolumn{2}{c}{\name} & \multicolumn{2}{c}{X-ray only} \\
\cmidrule(lr){2-3} \cmidrule(lr){4-5}
Parameter & $\hat R$ & ESS & $\hat R$ & ESS \\
\midrule
$\xi$       & 1.000 & 726,249 & 1.000 & 766,922 \\
$\psi$      & 1.004 & 4,367   & 1.000 & 22,615 \\
$\theta_0$  & 1.002 & 7,911   & 1.000 & 29,285 \\
$\theta_1$  & 1.000 & 384,115 & 1.000 & 572,631 \\
$\theta_2$  & 1.000 & 623,794 & 1.000 & 639,901 \\
$\tau_0$    & 1.009 & 512     & 1.000 & 26,604 \\
$\tau_1$    & 1.000 & 141,193 & 1.000 & 133,490 \\
$\tau_2$    & 1.000 & 742,209 & 1.000 & 756,652 \\
$\phi_{21}$ & 1.002 & 10,876  & --    & -- \\
$\phi_{22}$ & 1.001 & 3,825   & --    & -- \\
\bottomrule
\end{tabular}
\end{table}

\begin{table}[h]
\centering
\caption{MCMC diagnostics for the double-Pareto CDF-S analysis. Reported are $\hat R$ and effective sample size (ESS) for the \name\ and X-ray-only fits. The \name\ analysis used eight chains, while the X-ray-only analysis used four chains.}
\label{tab:double_pareto_diagnostics}
\begin{tabular}{lrrrr}
\toprule
& \multicolumn{2}{c}{\name} & \multicolumn{2}{c}{X-ray only} \\
\cmidrule(lr){2-3} \cmidrule(lr){4-5}
Parameter & $\hat R$ & ESS & $\hat R$ & ESS \\
\midrule
$\xi$       & 1.000 & 697,301 & 1.000 & 390,844 \\
$\psi$      & 1.009 & 3,591   & 1.001 & 18,132 \\
$\theta_1$  & 1.001 & 25,289  & 1.000 & 185,763 \\
$\theta_2$  & 1.000 & 626,968 & 1.000 & 318,269 \\
$\tau_1$    & 1.027 & 577     & 1.000 & 22,130 \\
$\tau_2$    & 1.000 & 777,498 & 1.000 & 378,771 \\
$\phi_{21}$ & 1.001 & 25,775  & --    & -- \\
$\phi_{22}$ & 1.010 & 8,334   & --    & -- \\
\bottomrule
\end{tabular}
\end{table}

For the double-Pareto analysis, most $\hat R$ values are also close to one.
The lower-bound parameter $\tau_1$ in the \name\ fit mixes more slowly, with $\hat R=1.027$ and an ESS of $577$, while the remaining parameters have substantially larger effective sample sizes. Because the X-ray-only double-Pareto fit used four chains rather than eight, its ESS values are based on fewer posterior draws and should not be compared directly with those from the \name\ fit. 

\newpage

\end{document}